\documentstyle[aps,eqsecnum,epsfig]{revtex}

\begin{document}
\draft
\title{Perturbative Stability of the QCD Predictions for\\
Single Spin Asymmetry in Heavy Quark Photoproduction\thanks{
Preprint LPT ORSAY 01-38}}
\author{\large N.Ya.Ivanov\footnote{E-mail: nikiv@uniphi.yerphi.am}}
\address{Yerevan Physics Institute, Alikhanian Br.2, 375036 Yerevan, 
Armenia\\}
\maketitle

\begin{abstract}
\noindent We present the threshold resummation for the single spin 
asymmetry in heavy flavor production by linearly polarized photons. 
We analyze the soft-gluon contributions to the asymmetry at fixed order 
in $\alpha_s$ to the next-to-leading logarithmic accuracy. 
Our analysis shows that, contrary to the production cross section, 
the azimuthal asymmetry is practically insensitive to soft radiation. 
Our calculations of the asymmetry up to the 6th order in 
$\alpha_s$ lead only to small corrections (of order of few percent) 
to the Born predictions at energies of the fixed target experiments. 
Fast convergence of the perturbative series for the azimuthal asymmetry 
is due to the factorization properties of the photon-hadron cross section. 
We conclude that measurements of the single spin asymmetry would provide 
an excellent test of pQCD applicability to heavy flavor production. 
\end{abstract}


\section{ Introduction}

\noindent Presently, the basic spin-averaged characteristics of heavy 
flavor hadro-, photo- and electroproduction are known exactly up to 
the next-to-leading order (NLO) (see \cite{1} for a review). 
Two main results of the explicit pQCD
calculations can be formulated as follows. First, the NLO corrections 
are large; they increase the leading order (LO) predictions for both 
charm and bottom production cross sections approximately by a factor 
of 2. For this reason, one could expect that the higher order corrections 
as well as the nonperturbative contributions can be essential in these 
processes, especially for the $c$-quark case. Second, the one-loop 
predictions are very sensitive to standard uncertainties in the input 
QCD parameters. In fact, the total uncertainties associated with the 
unknown values of the heavy quark mass, $m$, the factorization and 
renormalization scales, $\mu_{F}$ and $\mu _{R}$, $\Lambda _{QCD}$ and 
the parton distribution functions are so large that one can only 
estimate the order of magnitude of the NLO predictions for total 
cross sections \cite{2,3}. For this reason, it is very
difficult to compare directly, without additional assumptions, the NLO
predictions for spin-averaged cross sections with experimental data
and thereby to test the pQCD applicability to heavy quark production.

During the recent years, the role of higher order corrections has been 
extensively investigated in the framework of the soft gluon resummation 
formalism. For a review see Ref.\cite{4}.
Soft gluon (or threshold) resummation is based on the factorization 
properties of the cross section near the partonic threshold and makes it 
possible to resum to all orders in $\alpha_{s}$ the leading (Sudakov 
double) logarithms (LL) and the next-to-leading ones (NLL) \cite{5,6,7}.
Formally resummed cross sections are ill-defined due to the Landau pole 
contribution, and some prescription must be implemented to avoid the 
renormalon ambiguities. To obtain numerical predictions for the physical 
cross sections, a few prescriptions have been proposed \cite{8,9,10}.
Physically, the choice of a resummation prescription implies the 
introduction of an effective scale, separating the hard (perturbative) 
and soft (nonperturbative) gluon contributions. 
Another open question, also closely related to the convergence of the 
perturbative series, is the role of the subleading logarithms which are not, 
in principle, under control of the resummation procedure \cite{11,12}. 
Numerically, the higher order corrections to the heavy quark production 
cross sections can depend significantly on the choice of resummation 
prescription \cite{12}.

Since the spin-averaged characteristics of heavy flavor production are not
well defined quantitatively in pQCD it is of special interest to study those
spin-dependent observables which are stable under variations of input
parameters of the theory \cite{13}. In this paper we analyze the charm and
bottom production by linearly polarized photons, namely the reactions
\begin{equation}
\gamma \uparrow +h\rightarrow Q+X[\overline{Q}].  \label{1}
\end{equation}

We consider the single spin asymmetry parameter, $A(S)$, which measures the
parallel-perpendicular asymmetry in the quark azimuthal distribution:
\begin{equation}
\frac{\text{d}\sigma _{\gamma h}}{\text{d}\varphi }(S,\varphi )=\frac{\sigma
_{\gamma h}^{\text{{\rm unp}}}(S)}{2\pi }(1+A(S){\cal P}_{\gamma }\cos
2\varphi ),  \label{2}
\end{equation}
where
\begin{equation}
A(S)=\frac{1}{{\cal P}_{\gamma }}\frac{\text{d}\sigma _{\gamma h}(S,\varphi
=0)-\text{d}\sigma _{\gamma h}(S,\varphi =\pi /2)}{\text{d}\sigma _{\gamma
h}(S,\varphi =0)+\text{d}\sigma _{\gamma h}(S,\varphi =\pi /2)}.  \label{3}
\end{equation}
Here $\sigma_{\gamma h}^{\text{{\rm unp}}}(S)$ is the unpolarized cross 
section, d$\sigma (S,\varphi )\equiv \frac{\text{d}\sigma }{\text{d}\varphi}
(S,\varphi )$ , ${\cal P}_{\gamma }$ is the degree of linear polarization of
the incident photon beam, $\sqrt{S}$ is the centre of mass energy of the
process (\ref{1}) and $\varphi $ is the angle between the beam polarization
direction and the observed quark transverse momentum.

The Born level predictions  for the single spin asymmetry in (\ref{1})  as well 
as the contributions of nonperturbative effects (such as the gluon transverse
motion in the target and the heavy quark fragmentation) have been considered
in \cite{13}. The following remarkable properties of $A(S)$ have been 
observed:
\begin{itemize}
\item  The azimuthal asymmetry (\ref{3})  is of leading twist; in a wide 
region of initial energy, it is predicted to be about $20\%$ for both charm 
and bottom quark production.
\item  At energies sufficiently above the production threshold, the LO 
predictions for $A(S)$ are insensitive (to within few percent) to uncertainties 
in the QCD input parameters: $m$, $\mu _{R}$, $\mu _{F}$, $\Lambda _{QCD}$ and 
in the gluon distribution function. This implies that theoretical uncertainties 
in the spin-dependent and spin-averaged cross sections (the numerator and 
denominator of the fraction (\ref{3}), respectively) cancel each other with a 
good accuracy.
\item  Nonperturbative corrections to the $b$-quark azimuthal asymmetry $A(S)$ 
due to the gluon transverse motion in the target are negligible. Because
of the smallness of the $c$-quark mass, the analogous corrections to $A(s)$
in the charm case are larger; they are of the order of 20\%.
\end{itemize}

In the present paper, the soft-gluon corrections to the asymmetry (\ref{3}) 
are investigated. Using the methods of Refs.\cite{5,6,7} for the resummation 
of soft gluons, we resum the Sudakov logarithms for the polarized cross 
section (\ref{2}) in moment space to the NLL accuracy. Considering the obtained 
expression as a generating functional of the perturbative theory, we re-expand 
it in $\alpha_{s}$ in momentum space and analyze the fixed order predictions 
for the azimuthal asymmetry.
Our main results can be formulated as follows:
\begin{itemize}
\item  Contrary to the production cross sections, the asymmerty (\ref{3}) 
in azimuthal distributions of both charm and bottom quark is
practically insensitive to radiative corrections at fixed target energies. 
This implies that large soft-gluon contributions to the spin-dependent and 
unpolarized cross sections cancel each other in (\ref{3}) with a good accuracy.
\item  At the NLL level, the NLO and NNLO predictions for $A(S)$ affect the 
LO results by less than 1\% and 2\%, respectively.
\item  Our computations of higher order corrections indicate a fast convergence 
of the perturbative series for $A(S)$. We have calculated the asymmetry 
parameter $A(S)$ up to the 6th order (N$^{6}$LO) in $\alpha_{s}$ to the NNL 
accuracy and have found that corresponding corrections to the Born level 
result are of order of few percent.
\end{itemize}

Our analysis shows that high perturbative stability of $A(S)$ 
is due to the factorization properties of the photon-hadron cross section.
We conclude that, in contrast with the production cross sections, the single 
spin asymmetry in heavy flavor photoproduction is an observable quantitatively 
well defined in pQCD: it is stable, both parametricaly and perturbatively, and 
insensitive to nonperturbative corrections. 
Measurements of the azimuthal asymmetry in bottom photoproduction would 
provide an ideal test of the conventional parton model based on pQCD. 

Concerning the experimental aspects, the azimuthal asymmetry in charm 
photoproduction can be measured at SLAC where a coherent bremsstrahlung 
beam of linearly polarized photons with energies up to 40 GeV will be  
available soon \cite{15,16}.
Due to the $c$-quark low mass,  data on the $D$-meson azimuthal distributions 
would make it possible to clarify the role of subleading twist contributions \cite{13}.

The paper is organized as follows. Section II contains the derivation of the 
resummed formula for the spin-dependent photoproduction of heavy flavor in the 
single-particle inclusive kinematics. In Section III we analyze the NLO and 
NNLO predictions for the azimuthal asymmetry. We check the quality of the NNL 
approximation against available explicit results and discuss the subleading 
logarithms contribution. 
Higher order corrections and the role of the gluon distribution function in 
perturbative stability of the asymmetry are considered in Section IV.

\section{Soft gluon resummation for polarized cross section}

\noindent In this Section we carry out the resummation of the Sudakov
logarithms for the spin-dependent cross section of the reaction
\begin{equation}
\gamma (k_{\gamma })\uparrow +h(k_{h})\rightarrow Q(p_{Q})+X[\overline{Q}
](p_{X})  \label{4}
\end{equation}
to next-to-leading logarithmic (NLL) accuracy to all orders of the
perturbative expansion. In the single-particle inclusive (1PI) kinematics,
the overall invariants are defined as
\begin{eqnarray}
S &=&\left( k_{\gamma }+k_{h}\right) ^{2};\text{ \ \hspace{0.1in}\quad
\qquad ~ \hspace{0.1in} \thinspace \thinspace \hspace{0.05in}}T_{1}=\left(
k_{h}-p_{Q}\right) ^{2}-m^{2};  \nonumber \\
S_{4} &=&S+T_{1}+U_{1};\text{ \ \hspace{0.1in} \quad \qquad \thinspace ~ ~\ }
U_{1}=\left( k_{\gamma }-p_{Q}\right) ^{2}-m^{2}.  \label{5}
\end{eqnarray}
At the parton level, the dominant subprocess is the photon-gluon fusion:
\begin{equation}
\gamma (k_{\gamma })+g(k_{g})\rightarrow Q(p_{Q})+X[\overline{Q}](p_{X}),
\label{6}
\end{equation}
where the corresponding kinematical variables are
\begin{eqnarray}
s &=&\left( k_{\gamma }+k_{g}\right) ^{2}=zS;\text{ \ \hspace{0.1in} \ ~ 
\hspace{0.1in} \hspace{0.05in}}t_{1}=\left( k_{g}-p_{Q}\right)
^{2}-m^{2}=zT_{1};  \nonumber \\
s_{4} &=&s+t_{1}+u_{1};\text{ \ \hspace{0.1in} ~ \hspace{0.1in} \hspace{
0.05in} \ \quad ~~}u_{1}=U_{1},  \label{7}
\end{eqnarray}
with $k_{g}=zk_{h}$, $m$ the heavy quark mass, while $s_{4}$ measures the
inelasticity of the partonic reaction. At the Born level, ${\cal O}(\alpha
_{em}\alpha _{s})$, the only partonic subprocess which is responsible for
heavy quark photoproduction is the two-body photon-gluon fusion:
\begin{equation}
\gamma (k_{\gamma })+g(k_{g})\rightarrow Q(p_{Q})+\overline{Q}(p_{\stackrel{
\_}{Q}}).  \label{8}
\end{equation}
To take into account the NLO contributions, one needs to calculate the 
virtual ${\cal O}(\alpha _{em}\alpha _{s}^{2})$ corrections to the Born 
process (\ref{8}) and the real gluon emission:
\begin{equation}
\gamma (k_{\gamma })+g(k_{g})\rightarrow Q(p_{Q})+\overline{Q}(p_{\stackrel{
\_}{Q}})+g(p_{g}).  \label{9}
\end{equation}
We neglect the photon-(anti)quark fusion subprocesses as well as the 
so-called hadronic or resolved component of the photon. 
This is justified as their contributions vanish at LO and are small at 
NLO at the energies under consideration \cite{1,2}.

The factorized single-heavy quark inclusive cross section for photon-hadron
collisions, d$\sigma _{\gamma h}$, has the form of a convolution of the
perturbative short-distance cross section, d$\hat{\sigma}_{\gamma g}$, with
the universal parton distribution function $\phi _{g/h}$:
\begin{equation}
\text{d}\sigma _{\gamma h}\left( S,T_{1},U_{1},\varphi ,\mu _{F},\alpha
_{s}(\mu _{R}^{2})\right) =\int \text{d}z\phi _{g/h}(z,\mu _{F})\text{d}\hat{
\sigma}_{\gamma g}\left( s,t_{1},u_{1},\varphi ,\mu _{F},\alpha _{s}(\mu
_{R}^{2})\right) ,  \label{10}
\end{equation}
where $\mu _{F}$ and $\mu _{R}$ are the factorization and renormalization
scales, respectively. Note that here and in the following d$\sigma _{\gamma
h}$ and d$\hat{\sigma}_{\gamma g}$ denote any relevant $\varphi $-dependent
differential distribution. Replacing the incoming hadron by the gluon and
taking the Laplace moments, the above convolution simplifies to a product:
\begin{equation}
\text{d}\tilde{\sigma}_{\gamma g}(N,\varphi )=\tilde{\phi}_{g/g}(N_{u})\text{
d}\hat{\sigma}_{\gamma g}(N,\varphi ).  \label{11}
\end{equation}
The moments for d$\hat{\sigma}_{\gamma g}$ are defined by
\begin{equation}
\text{d}\hat{\sigma}_{\gamma g}(N,\varphi )=\int_{0}^{\infty }\frac{\text{d}
s_{4}}{m^{2}}\text{e}^{-Ns_{4}/m^{2}}\text{d}\hat{\sigma}_{\gamma
g}(s_{4},\varphi ),  \label{12}
\end{equation}
with $N$ the moment variable. The upper limit of this integral is not
important for large $N$ and may be put at $1$. Similarly, the moments 
for $\phi _{g/g}$ with respect to $z$ have the form
\begin{equation}
\tilde{\phi}_{g/g}(N_{u})=\int_{0}^{1}\text{d}z\text{e}^{-N_{u}z}\phi
_{g/g}(z),\quad \quad N_{u}=N(-u_{1}/m^{2}),  \label{13}
\end{equation}
where definition of $N_{u}$ is given for the 1PI kinematics \cite{6,17}.

The short-distance perturbative cross section d$\hat{\sigma}_{\gamma g}$
still be sensitive to the collinear gluon emission, $\vec{p}
_{g,T}\rightarrow 0$. To separate these collinear effects from the hard
scattering, a refactorization is introduced \cite{18,5}:

\begin{equation}
\text{d}\tilde{\sigma}_{\gamma g}(N,\varphi )=\tilde{\psi}
_{g/g}(N_{u})H_{\gamma g}(\varphi )\tilde{S}_{\gamma g}\left( \frac{m}{N\mu
_{F}}\right) ,  \label{14}
\end{equation}
where $\psi _{g/g}$ is the center-of-mass parton distribution that absorb
the universal collinear singularities associated with the initial-state
gluon while $\tilde{S}_{\gamma g}$ is the soft-gluon function that describes
the non-collinear soft gluon emission. The mass of the heavy quarks protects 
the final state from collinear singularities. The hard-scattering part of 
cross section, $H_{\gamma g}$, is free of soft-gluon effects and thus 
independent of $N$. 
In the photoproduction case, both $H_{\gamma g}$ and $\tilde{S}_{\gamma g}$ 
are simply functions, and not matrices in color space, in contrast with heavy 
quark or jet production in hadron-hadron collisions \cite{7}.

Comparing Eqs. (\ref{11}) and (\ref{14}), we derive
\begin{equation}
\text{d}\hat{\sigma}_{\gamma g}(N,\varphi )=\frac{\tilde{\psi}_{g/g}(N_{u})}{
\tilde{\phi}_{g/g}(N_{u})}\tilde{S}_{\gamma g}\left( \frac{m}{N\mu _{F}}
\right) H_{\gamma g}(\varphi ).  \label{15}
\end{equation}
The functions $\tilde{\psi}_{g/g}/\tilde{\phi}_{g/g}$ and $\tilde{S}_{\gamma
g}$ originate from the collinear, $\vec{p}_{g,T}\rightarrow 0$, and soft, $
\vec{p}_{g}\rightarrow 0$, limits. Since the azimuthal angle $\varphi $ is
the same for both $\gamma g$ and $Q\overline{Q}$ center-of-mass systems in
these limits, only the hard function, $H_{\gamma g}$, is $\varphi $-
dependent in the right-hand part of (\ref{15}).

In the {\large $\overline{\text{MS}}$ }factorization scheme, at NLO and to
NLL accuracy, the ratio $\tilde{\psi}_{g/g}/\tilde{\phi}_{g/g}$ is \cite{17}
\begin{equation}
\frac{\tilde{\psi}_{g/g}}{\tilde{\phi}_{g/g}}\left( N_{u},\mu _{F}\right) =1+
\frac{\alpha _{s}C_{A}}{\pi }\left[ \ln ^{2}\tilde{N}_{u}+\left( 1+\ln \frac{
\mu _{F}^{2}}{m^{2}}\right) \ln \tilde{N}_{u}\right] ,  \label{16}
\end{equation}
where $\tilde{N}_{u}=N_{u}$e$^{\gamma _{E}}$ with $\gamma _{E}$ the Euler
constant. The function $\tilde{\psi}_{g/g}/\tilde{\phi}_{g/g}$ summarizes
all leading logarithms of $\tilde{N}$ and a part of the next-to-leading
ones. (At NLO, they are $\ln ^{2}\tilde{N}$ and $\ln \tilde{N}$,
respectively). $\tilde{S}_{\gamma g}$-function contributes only at the
next-to-leading level of $\ln \tilde{N}$.

To resum the leading, ${\cal O}\left( \alpha _{s}^{n}\ln ^{2n}\tilde{N}
\right) $, and next-to-leading, ${\cal O}\left( \alpha _{s}^{n}\ln ^{2n-1}
\tilde{N}\right) $, logarithms for d$\hat{\sigma}_{\gamma g}(N,\varphi )$ 
to all orders of $n$, we use the methods of Refs. \cite{5,6,7} based on the 
solving of the appropriate evolution equations for each of the functions 
in (\ref{15}). The formal result for the resummed spin-dependent cross 
section in moment space is:
\begin{eqnarray}
\text{d}\hat{\sigma}_{\gamma g}(N,t_{1},u_{1},\varphi ,\mu _{F},\alpha
_{s}(\mu _{R}^{2})) &=&\exp \left[ E_{g}(N_{u})\right] H_{\gamma
g}(t_{1},u_{1},\varphi ,\alpha _{s}(\mu _{R}^{2}))\tilde{S}_{\gamma g}\left(
1,\alpha _{s}\left( m^{2}/N^{2}\right) \right) \times   \nonumber \\
&&\exp \left[ \int_{m}^{m/N}\frac{\text{d}\mu ^{\prime }}{\mu ^{\prime }}2\,
\text{Re}\Gamma _{S}^{\gamma g}\left( \alpha _{s}\left( \mu ^{\prime
2}\right) \right) \right] .  \label{17}
\end{eqnarray}
The first exponent in (\ref{17}) resums the $N$-dependence of the ratio 
$\tilde{\psi}_{g/g}/\tilde{\phi}_{g/g}$ and is given in the 
{\large $\overline{\text{MS}}$ }scheme by
\begin{equation}
E_{g}(N_{u})=\int_{0}^{\infty }\frac{\text{d}\omega }{\omega }\left( 1-\text{
e}^{-N_{u}\omega }\right) \left[ \int_{\omega ^{2}m^{2}}^{\mu _{F}^{2}}\frac{
\text{d}\mu ^{\prime 2}}{\mu ^{\prime 2}}A_{g}\left( \alpha _{s}\left( \mu
^{\prime 2}\right) \right) +\frac{1}{2}\nu _{g}\left( \alpha _{s}\left(
\omega ^{2}m^{2}\right) \right) \right] .  \label{18}
\end{equation}
The function $A_{g}$ is known at two loops \cite{18,19}, $A_{g}\left(\alpha _{s}
\right)=\alpha _{s}C_{A}/\pi +\left( \alpha _{s}/\pi \right) ^{2}C_{A}K/2$, 
$K=C_{A}\left( 67/18-\pi ^{2}/6\right) -5n_{f}/9$, where $n_{f}$ is the
number of active quark flavors, and $\nu _{g}=2\alpha _{s}C_{A}/\pi$ \cite{7}.

The second exponent controls the evolution of the soft function from scale 
$m/N$ to $m$ and is given in terms of $\Gamma _{S}^{\gamma g}$, the soft
anomalous dimension. $\Gamma _{S}^{\gamma g}$ was calculated at one loop in
\cite{17}:

\begin{equation}
\Gamma _{S}^{\gamma g}(\alpha _{s})=\frac{\alpha _{s}C_{A}}{2\pi }\left[ \ln 
\frac{u_{1}t_{1}}{m^{4}}+\left( 1-\frac{2C_{F}}{C_{A}}\right) \left(
1+L_{\beta }\right) \right] .  \label{19}
\end{equation}
Here

\begin{equation}
L_{\beta }=\frac{1-2m^{2}/s}{\beta }\left[ \text{ln}\left( \frac{1-\beta }{
1+\beta }\right) +\text{i}\pi \right] ,\qquad \quad \beta =\sqrt{1-4m^{2}/s},
\label{20}
\end{equation}
while $C_{A}=N_{c}$ and $C_{F}=(N_{c}^{2}-1)/(2N_{c})$, where $N_{c}$ is the
number of colors. The product $H_{\gamma g}\tilde{S}_{\gamma g}$ on the
first line of Eq.(\ref{17}) is determined from matching at lowest order 
in $\alpha _{s}$ to the Born result.

The exponent in Eqs.(\ref{17}), (\ref{18}) can only be interpreted in a formal 
sense since the corresponding integration over $\omega $ near $\Lambda _{QCD}/m$ 
is not defined unambiguously due to the Landau pole in the coupling strength 
$\alpha _{s}$. To avoid this soft gluon divergence, several prescriptions have 
been proposed \cite{8,9,10}. However, if one expands the exponents in the 
resummed cross section at fixed order in $\alpha _{s}$, no divergences 
associated with the Landau pole are encountered.

\section{NLO and NNLO predictions}
\subsection{Partonic Cross Sections}

\noindent In this section we expand the resummed cross section (\ref{17}) 
to one and two loop order and invert back to momentum space using the equation 
\cite{20}
\begin{equation}
\left( -\ln \tilde{N}\right) ^{l+1}=(l+1)\int_{0}^{\infty }\text{d}s_{4}
\text{e}^{-Ns_{4}/m^{2}}\left[ \frac{\ln ^{l}\left( s_{4}/m^{2}\right) }{
s_{4}}\right] _{+}+{\cal O}\left( \ln ^{l-1}\tilde{N}\right) ,  \label{21}
\end{equation}
where 1PI singular ``plus`` distributions are defined by
\begin{equation}
\left[ \frac{\ln ^{l}\left( s_{4}/m^{2}\right) }{s_{4}}\right]
_{+}=\lim_{\epsilon \rightarrow 0}\left\{ \frac{\ln ^{l}\left(
s_{4}/m^{2}\right) }{s_{4}}\theta \left( s_{4}-\epsilon \right) +\frac{1}{l+1}
\ln ^{l+1}\left( \frac{\epsilon }{m^{2}}\right) \delta \left( s_{4}\right)
\right\} .  \label{22}
\end{equation}
For any sufficiently regular test function $h(s_{4})$,  Eq.(\ref{22}) gives
\begin{equation}
\int_{0}^{s_{4}^{\max }}\text{d}s_{4}h(s_{4})\left[ \frac{\ln ^{l}\left(
s_{4}/m^{2}\right) }{s_{4}}\right] _{+}=\int_{0}^{s_{4}^{\max }}\text{d}
s_{4}\left[ h(s_{4})-h(0)\right] \frac{\ln ^{l}\left( s_{4}/m^{2}\right) }{
s_{4}}+\frac{1}{l+1}h(0)\ln ^{l+1}\left( s_{4}^{\max }/m^{2}\right).  \label{add3}
\end{equation}
Eq.(\ref{21}) provides us the NLL approximation for both spin-dependent, 
d$\Delta \hat{\sigma}_{\gamma g}$, and unpolarized, 
d$\hat{\sigma}_{\gamma g}$, differential cross sections:
\begin{equation}
s^{2}\frac{\text{d}^{3}\hat{\sigma}_{\gamma g}\left( s,t_{1},u_{1},\varphi
\right) }{\text{d}\varphi \text{d}t_{1}\text{d}u_{1}}= \frac{1}{2\pi }
\left( s^{2}\frac{\text{d}^{2}\hat{\sigma}_{\gamma g}\left(
s,t_{1},u_{1}\right) }{\text{d}t_{1}\text{d}u_{1}}+s^{2}\frac{\text{d}
^{2}\Delta \hat{\sigma}_{\gamma g}\left( s,t_{1},u_{1}\right) }{\text{d}t_{1}
\text{d}u_{1}}{\cal P}_{\gamma }\cos 2\varphi \right) ,  \label{23}
\end{equation}
where ${\cal P}_{\gamma }$ is the degree of the photon beam polarization; $
\varphi $ is the angle between the observed quark transverse momentum, $\vec{
p}_{Q,T}$, and the beam polarization direction.

To NLL accuracy, the perturbative expansion for the partonic cross section can
be written in a factorized form as

\begin{equation}
s^{2}\frac{\text{d}^{2}(\Delta )\hat{\sigma}_{\gamma g}\left(
s,t_{1},u_{1}\right) }{\text{d}t_{1}\text{d}u_{1}}=(\Delta )B_{\gamma g}^{
\text{{\rm Born}}}\left( s,t_{1},u_{1}\right) \left\{ \delta \left(
s+t_{1}+u_{1}\right) +\sum_{n=1}^{\infty }\left( \frac{\alpha _{s}C_{A}
}{\pi }\right) ^{n}K^{(n)}\left( s,t_{1},u_{1}\right) \right\} ,  \label{24}
\end{equation}
with the Born level hard parts $(\Delta )B_{\gamma g}^{\text{Born}}$ given
by \cite{21,13}
\begin{equation}
B_{\gamma g}^{\text{{\rm Born}}}\left( s,t_{1},u_{1}\right) =2\pi
e_{Q}^{2}\alpha _{em}\alpha _{s}T_{F}\left[ \frac{t_{1}}{u_{1}}+\frac{u_{1}}{
t_{1}}+\frac{4m^{2}s}{t_{1}u_{1}}\left( 1-\frac{m^{2}s}{t_{1}u_{1}}\right)
\right] ,  \label{25}
\end{equation}
\begin{equation}
\Delta B_{\gamma g}^{\text{{\rm Born}}}\left( s,t_{1},u_{1}\right) =2\pi
e_{Q}^{2}\alpha _{em}\alpha _{s}T_{F}\left[ \frac{4m^{2}s}{t_{1}u_{1}}\left(
1-\frac{m^{2}s}{t_{1}u_{1}}\right) \right] .  \label{26}
\end{equation}
In (\ref{25}) and (\ref{26}), $T_{F}=$Tr$(T^{a}T^{a})/(N_{c}^{2}-1)=1/2$ is 
the color factor, and $e_{Q}$ is the quark charge in units of electromagnetic 
coupling constant.

The NLO and NNLO soft gluon corrections to NLL accuracy are
\begin{eqnarray}
K^{(1)}\left( s,t_{1},u_{1}\right)  &=&2\left[ \frac{\ln \left(
s_{4}/m^{2}\right) }{s_{4}}\right] _{+}-\left[ \frac{1}{s_{4}}\right]
_{+}\left\{ 1+\ln \left( \frac{u_{1}}{t_{1}}\right) -\left( 1-\frac{2C_{F}}{
C_{A}}\right) \left( 1+\text{Re}L_{\beta }\right) +\ln \left( \frac{\mu ^{2}
}{m^{2}}\right) \right\} +  \nonumber \\
&&\delta \left( s_{4}\right) \ln \left( \frac{-u_{1}}{m^{2}}\right) \ln
\left( \frac{\mu ^{2}}{m^{2}}\right) ,  \label{27}
\end{eqnarray}
and
\begin{eqnarray}
K^{(2)}\left( s,t_{1},u_{1}\right)  &=&2\left[ \frac{\ln ^{3}\left(
s_{4}/m^{2}\right) }{s_{4}}\right] _{+}-  \nonumber \\
&&3\left[ \frac{\ln ^{2}\left( s_{4}/m^{2}\right) }{s_{4}}\right]
_{+}\left\{ 1+\ln \left( \frac{u_{1}}{t_{1}}\right) -\left( 1-\frac{2C_{F}}{
C_{A}}\right) \left( 1+\text{Re}L_{\beta }\right) +\frac{2}{3}\frac{b_{2}}{
C_{A}}+\ln \left( \frac{\mu ^{2}}{m^{2}}\right) \right\} +  \nonumber \\
&&2\left[ \frac{\ln \left( s_{4}/m^{2}\right) }{s_{4}}\right] _{+} \left\{ 1+
\ln \left( \frac{u_{1}}{t_{1}}
\right) -\left( 1-\frac{2C_{F}}{C_{A}}\right) \left( 1+\text{Re}L_{\beta
}\right) +\ln \left( \frac{-u_{1}}{m^{2}}\right) +\frac{b_{2}}{C_{A}}+\frac{1
}{2}\ln \left( \frac{\mu ^{2}}{m^{2}}\right) \right\}\times   \nonumber \\
&&\ln \left(\frac{\mu ^{2}}{m^{2}}\right)-\left[ \frac{1}{s_{4}}\right] _{+}
\ln ^{2}\left( \frac{\mu ^{2}}{m^{2}}
\right) \left\{ \ln \left( \frac{-u_{1}}{m^{2}}\right) +\frac{1}{2}\frac{
b_{2}}{C_{A}}\right\} ,  \label{28}
\end{eqnarray}
where we use $\mu =\mu _{F}=\mu _{R}$, and $b_{2}$ is the first coefficient of the 
$\beta \left( \alpha _{s}\right) $-function expansion: 
\begin{equation}
\beta \left( \alpha _{s}\right) \equiv \frac{\text{d}\ln \alpha _{s}\left(
\mu ^{2}\right) }{\text{d}\ln \mu ^{2}}=-\sum_{k=1}^{\infty }b_{k+1}(\alpha
_{s}/\pi )^{k},  \label{29}
\end{equation}
$b_{2}=\left( 11C_{A}-2n_{f}\right) /12$, 
$b_{3}=\left(34C_{A}^{2}-10C_{A}n_{f}-16C_{A}n_{f}\right) /48$. 
In (\ref{27}) and (\ref{28}), we have preserved the NLL terms for all the orders of 
the scale-dependent logarithm $\ln^{k}\left( \mu ^{2}/m^{2}\right) $, ($k=0,1,2$). 
We have checked that the result (\ref{27}) agrees to NLL accuracy with the exact 
${\cal O}(\alpha_{em}\alpha _{s}^{2})$ calculations of the unpolarized photon-gluon 
fusion given in \cite{23,24}, and that the Eqs.(\ref{27}) and (\ref{28}) coincide 
completely with the corresponding ones obtained in \cite{17,22} for the 
electroproduction case.

To perform a numerical investigation of the results (\ref{27}) and (\ref{28}), 
it is convenient to introduce for the fully inclusive (integrated over $t_{1}$ 
and $u_{1}$) cross sections, $\hat{\sigma}_{\gamma g}$ and 
$\Delta \hat{\sigma}_{\gamma g}$, 
\begin{equation}
(\Delta )\hat{\sigma} _{\gamma g}(s)=\int_{(1-\beta )s/2}^{(1+\beta )s/2}\text{d}
(-t_{1})\int_{0}^{s_{4}^{\max }}\text{d}s_{4}\frac{\text{d}^{2}(\Delta)
\hat{\sigma} _{\gamma g}}{\text{d}t_{1}\text{d}s_{4}}(s,t_{1},s_{4}), \label{add1}
\end{equation}
\begin{equation}
s_{4}^{\max }=s+t_{1}+\frac{m^{2}s}{t_{1}},  \label{add2}
\end{equation}
the dimensionless coefficient functions $(\Delta )c^{(k,l)}$,
\begin{equation}
(\Delta )\hat{\sigma}_{\gamma g}(\eta ,\mu ^{2})=\frac{\alpha _{em}\alpha
_{s}(\mu ^{2})e_{Q}^{2}}{m^{2}}\sum_{k=0}^{\infty }\left( 4\pi \alpha
_{s}(\mu ^{2})\right) ^{k}\sum_{l=0}^{k}(\Delta )c^{(k,l)}(\eta )\ln ^{l}
\frac{\mu ^{2}}{m^{2}},\qquad \qquad \eta =\frac{s}{4m^{2}}-1,  \label{30}
\end{equation}
where the variable $\eta $ measures the distance to the partonic threshold.

Concerning the scale-independent coefficient functions, only 
$c^{(1,0)}$ is known exactly \cite{25}. As to the $\mu $-dependent coefficients,
some of them can by calculated explicitly using the renormalization group
equation: 
\begin{equation}
\frac{\text{d}(\Delta )\hat{\sigma}_{\gamma g}^{\text{{\rm res}}}(s,\mu )}{
\text{d}\ln \mu ^{2}}=-\int_{\rho }^{1}\text{d}z(\Delta )\hat{\sigma}
_{\gamma g}^{\text{res}}(zs,\mu )P_{gg}(z),  \label{31}
\end{equation}
where $\rho =4m^{2}/s$, $(\Delta )\hat{\sigma}_{\gamma g}^{\text{{\rm res}}
}(s,\mu )$ are the cross sections resummed to all orders in $\alpha _{s}$
and $P_{gg}(z)$ is the corresponding (resummed) Altarelli-Parisi gluon-gluon
splitting function. Expanding Eq.(\ref{31}) in $\alpha _{s}$, we find
\begin{equation}
(\Delta )c^{(1,1)}=\frac{1}{4\pi ^{2}}\left[ b_{2}(\Delta )c^{(0,0)}-(\Delta
)c^{(0,0)}\otimes P_{gg}^{(0)}\right] ,  \label{32}
\end{equation}
\begin{equation}
(\Delta )c^{(2,1)}=\frac{1}{\left( 4\pi ^{2}\right) ^{2}}\left[ b_{3}(\Delta
)c^{(0,0)}-(\Delta )c^{(0,0)}\otimes P_{gg}^{(1)}\right] +\frac{1}{4\pi ^{2}}
\left[ 2b_{2}(\Delta )c^{(1,0)}-(\Delta )c^{(1,0)}\otimes
P_{gg}^{(0)}\right] ,  \label{33}
\end{equation}
\begin{eqnarray}
(\Delta )c^{(2,2)} &=&\frac{1}{4\pi ^{2}}\left[ b_{2}(\Delta )c^{(1,1)}-
\frac{1}{2}(\Delta )c^{(1,1)}\otimes P_{gg}^{(0)}\right] =  \label{34} \\
&&\frac{1}{\left( 4\pi ^{2}\right) ^{2}}\left[ b_{2}(\Delta )c^{(0,0)}-\frac{
3}{2}b_{2}(\Delta )c^{(0,0)}\otimes P_{gg}^{(0)}+\frac{1}{2}(\Delta
)c^{(0,0)}\otimes P_{gg}^{(0)}\otimes P_{gg}^{(0)}\right] ,  \nonumber
\end{eqnarray}
where the convolutions $(\Delta )c^{(k,l)}\otimes P_{gg}^{(j)}$ are defined
as
\begin{equation}
\left[ (\Delta )c^{(k,l)}\otimes P_{gg}^{(j)}\right] \left( s\right) \equiv
\int_{\rho }^{1}\text{d}z(\Delta )c^{(k,l)}(zs)P_{gg}^{(j)}(z),  \label{35}
\end{equation}
with the one- and two-loop gluon splitting functions, $P_{gg}(z)=(\alpha
_{s}/\pi )P_{gg}^{(0)}(z)+(\alpha _{s}/\pi )^{2}P_{gg}^{(1)}(z)+...$, given
in \cite{26}. Note that Eq.(\ref{32}) agrees with the exact result for $c^{(1,1)}$ 
given in \cite{25}.

With Eqs.(\ref{32})-(\ref{34}) in hand, we are able to check the quality of the 
NLL approximation at both NLO and NNLO against exact answers. 
In Figs.1 and 2 we plot the functions $c^{(k,l)}(\eta )$ and $\Delta
c^{(k,l)}(\eta )$, respectively. Predictions of the NLL approximation 
(\ref{27}), (\ref{28}) are given by dashed curves. The available exact results 
are given by solid lines. One can see a reasonable agreement up to energies 
$\eta \approx 2$. 

In Fig.3 we plot the NLL predictions for the ratios  
$\frac{\Delta c^{(n,0)}}{c^{(n,0)}}(\eta)$, $n=0,1,2$. The NNLO curve is given only up to 
$\eta=2$ since, at larger energies, the coefficient function $c^{(2,0)}$ changes sign and the 
ratio $\frac{\Delta c^{(2,0)}}{c^{(2,0)}}$ strongly oscillates. Fig.3 shows sizable deviations 
of the NLO and NNLO results from the Born level ones. This is due to the fact that the 
physical soft-gluon corrections (\ref{add1}) are determined by a convolution of the Born 
cross section with the Sudakov logarithms which, apart from  factorized 
$\delta(s_4)$-terms, contain also non-factorable ones (see Eq.(\ref{add3})).  
Kinematically, large values of $\eta$ allow  $s_{4}/m^{2}\gtrsim 1$ 
that leads to significant non-factorable  corrections. 
In other words,  the collinear bremsstrahlung carries away a large part of the initial energy. 
Since the spin-dependent and unpolarized Born level partonic cross sections have different 
energy behavior, the soft radiation has different impact on these quantities. 
\begin{figure}
\begin{center}
\begin{tabular}{cc}
\mbox{\epsfig{file=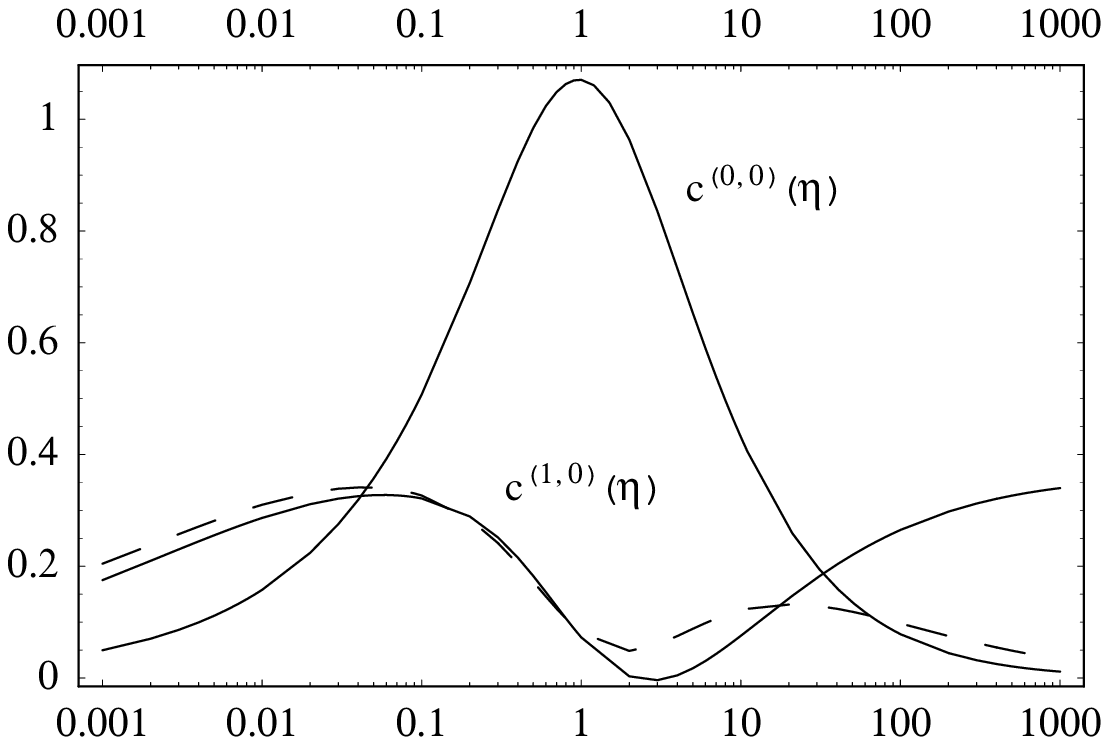,width=250pt}}
&\mbox{\epsfig{file=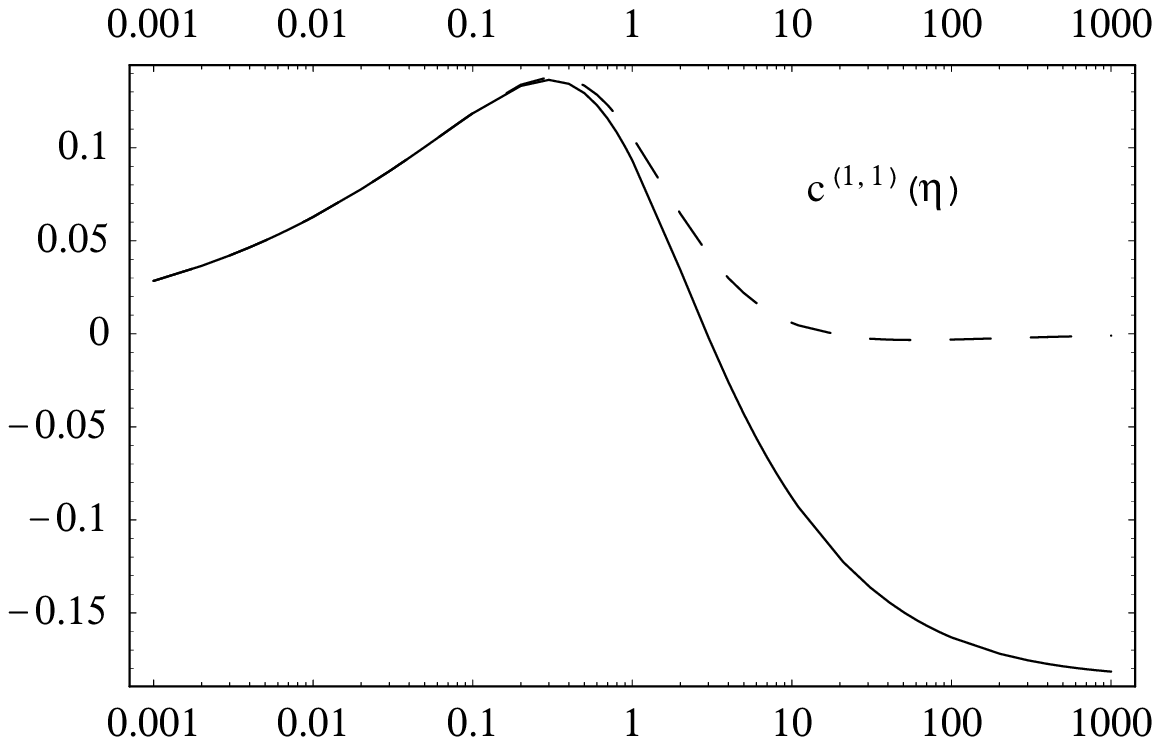,width=250pt}}\\
\mbox{\epsfig{file=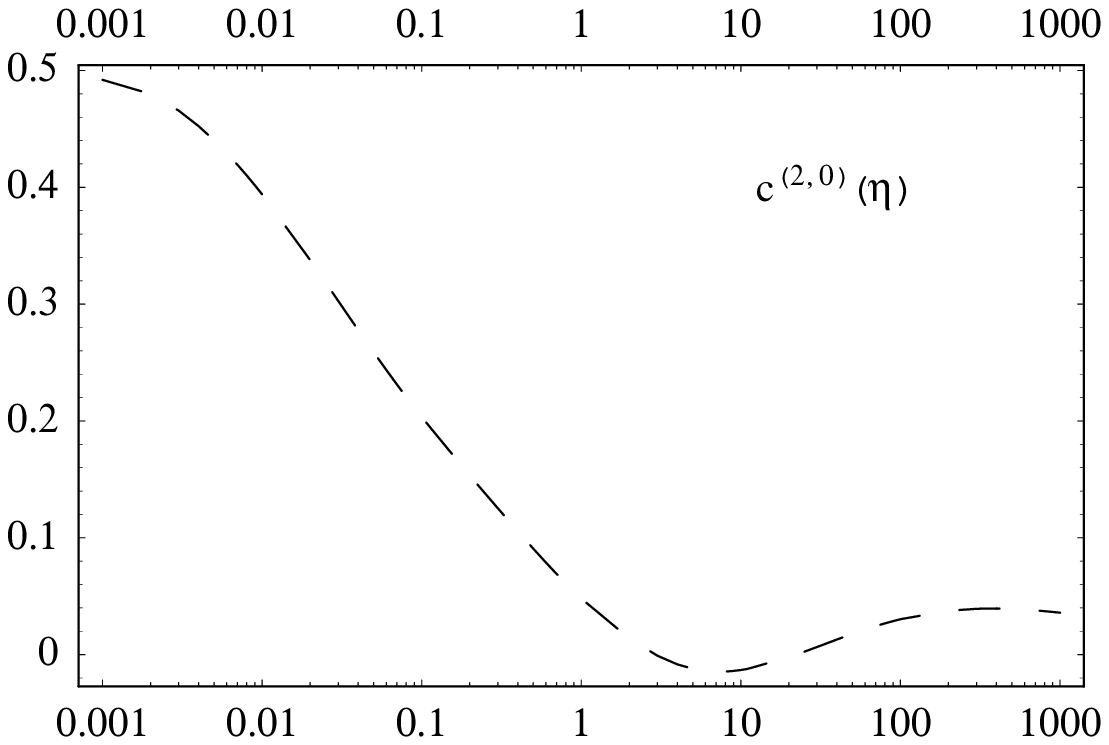,width=250pt}}
&\mbox{\epsfig{file=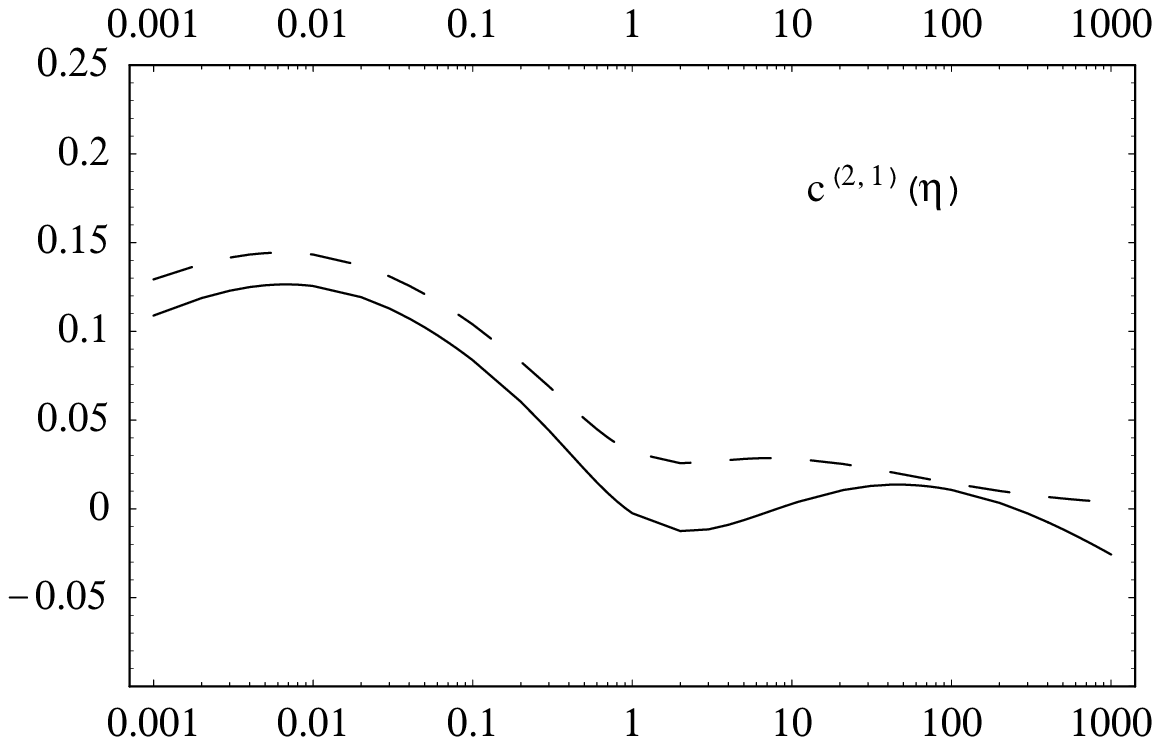,width=250pt}}\\
\multicolumn{2}{c}{\mbox{\epsfig{file=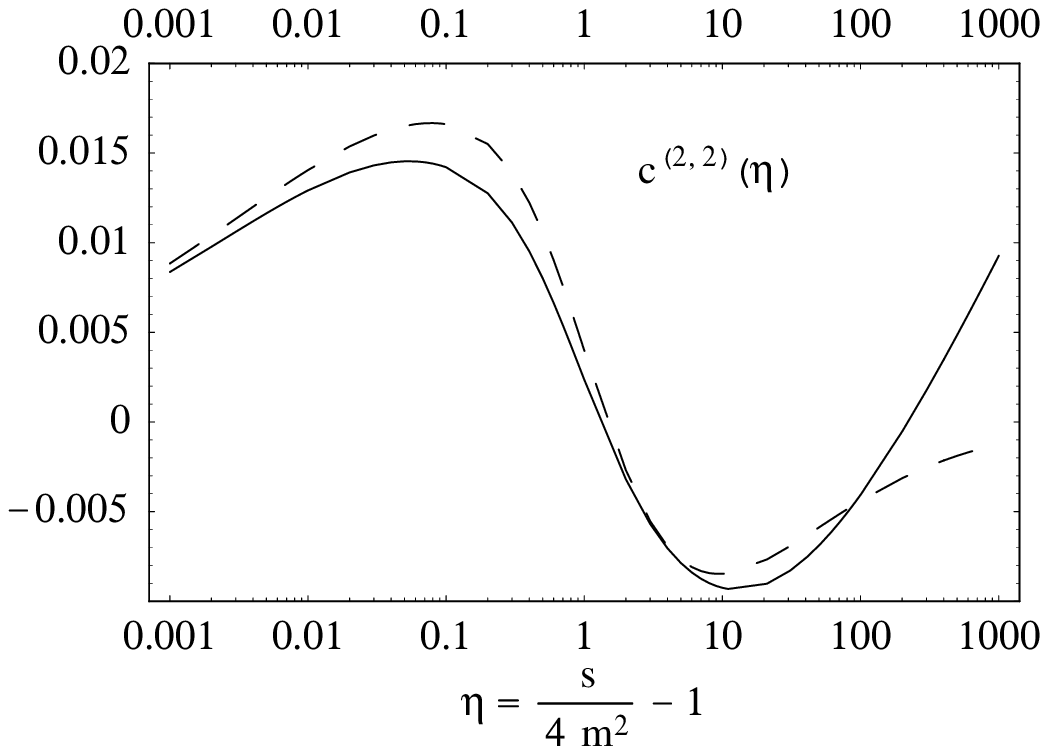,width=280pt}}}\\
\end{tabular}
\caption{\small $c^{(k,l)}(\eta )$ coefficient functions. Plotted are 
the available exact results (solid lines) and the NLL approximation 
(dashed lines).}
\end{center}
\end{figure}

\begin{figure}
\begin{center}
\begin{tabular}{cc}
\mbox{\epsfig{file=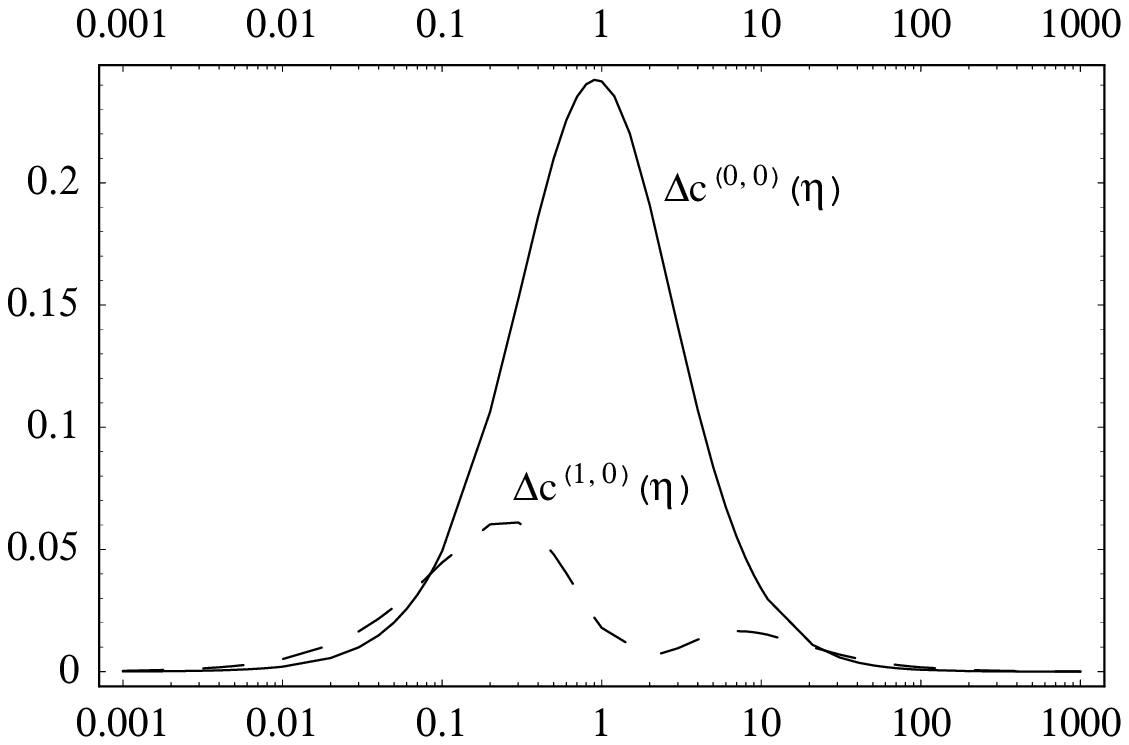,width=250pt}}
&\mbox{\epsfig{file=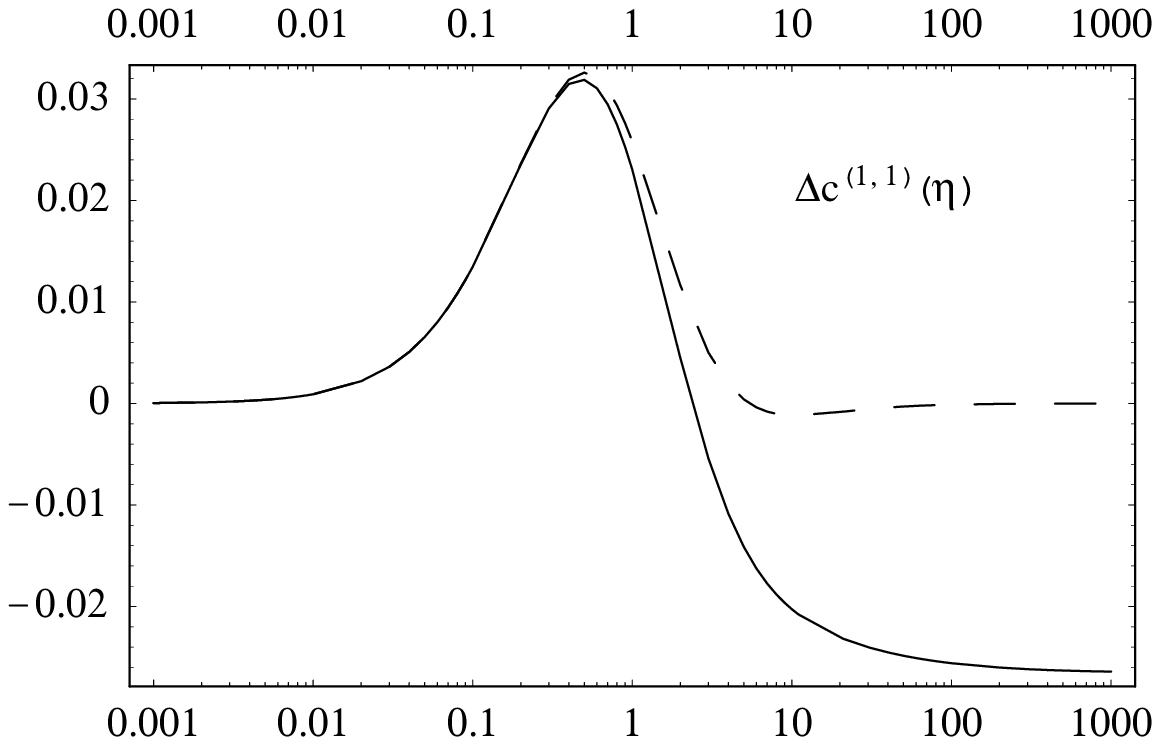,width=250pt}}\\
\mbox{\epsfig{file=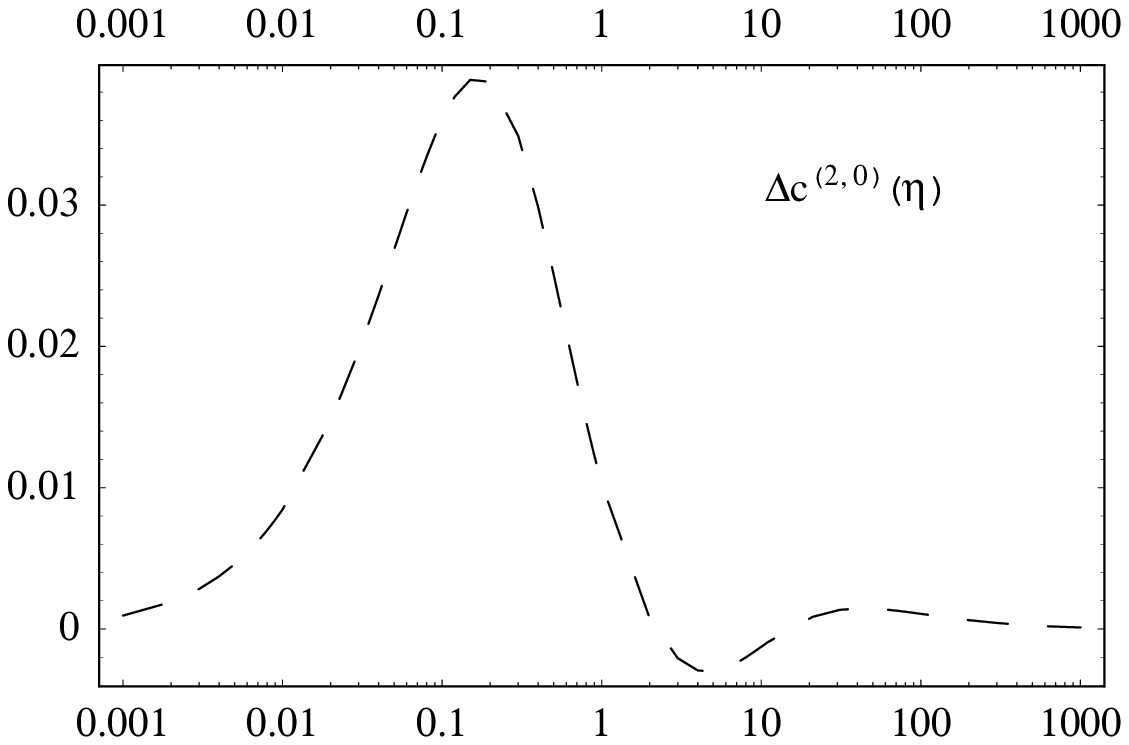,width=250pt}}
&\mbox{\epsfig{file=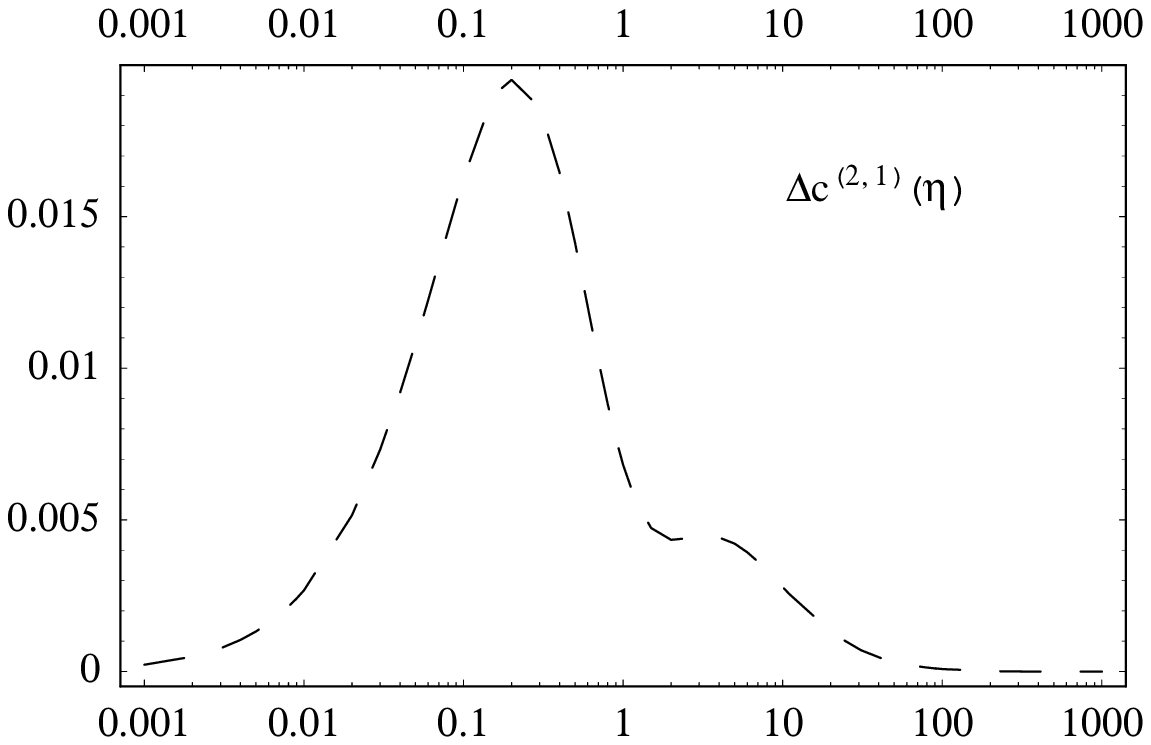,width=250pt}}\\
\multicolumn{2}{c}{\mbox{\epsfig{file=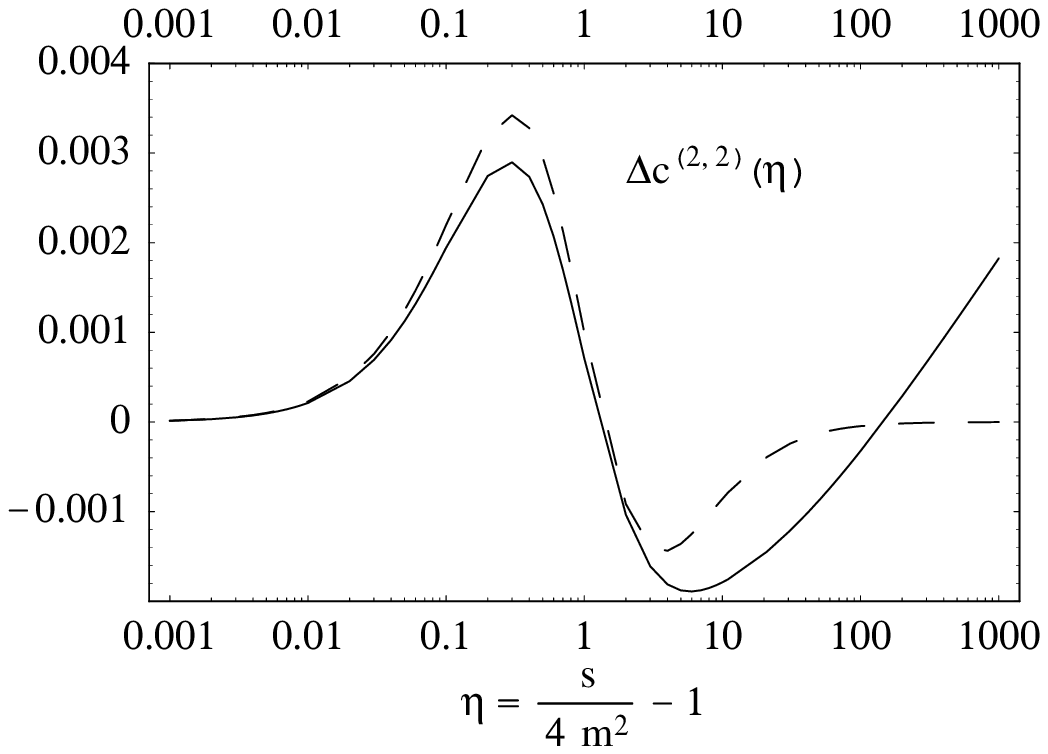,width=280pt}}}\\
\end{tabular}
\caption{\small $\Delta c^{(k,l)}(\eta )$ coefficient functions. Plotted are 
the available exact results (solid lines) and the NLL approximation 
(dashed lines).}
\end{center}
\end{figure}

\begin{figure}
\begin{center}
\begin{tabular}{cc}
\mbox{\epsfig{file=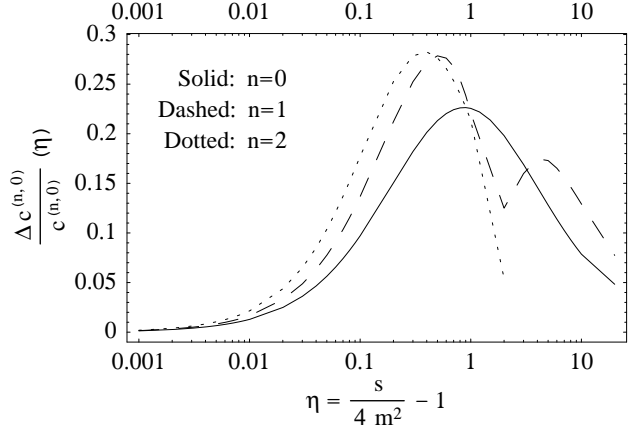,width=270pt}}\\
\end{tabular}
\caption{\small NLL predictions for the ratio 
$\frac{\Delta c^{(n,0)}}{c^{(n,0)}}(\eta)$. Plotted are the LO (solid curve), 
NLO (dashed curve) and NNLO (dotted curve) results.}
\end{center}
\end{figure}

\subsection{Hadron Level Results}

\noindent Let us now analyze the impact of the approximate NLO and NNLO perturbative
corrections on the azimuthal asymmetry, $A(S)$, at hadron level. Unless
otherwise stated, the CTEQ5M \cite{27} parametrization of the gluon distribution
function is used. The default values of the charm and bottom mass are 
$m_{c}= $ 1.5 GeV and $m_{b}=$ 4.75 GeV. For our analysis at NLO (NNLO) we
use the two-loop (three-loop \cite{28}) expression for $\alpha _{s}$; $\Lambda
_{4}= $ 300 MeV and $\Lambda _{5}=$ 200 MeV. The default values of the
factorization scale $\mu _{F}$ chosen for the $A(S)$ asymmetry calculation
are $\mu _{F}\mid _{{\rm {Charm}}}=2m_{c}$ for the case of charm production
and $\mu _{F}\mid _{{\rm {Bottom}}}=m_{b}$ for the bottom case \cite{1,29}. For 
the renormalization scale, $\mu _{R}$, we use $\mu _{R}=\mu _{F}$.

Our results for the single spin asymmetry $A(S)$ in charm and bottom
photoproduction at fixed target energies are presented in Fig.4. For
comparison, we plot in Fig.5 the so-called $K$-factors for unpolarized
cross sections: $K_{\gamma h}^{(1)}(S)=$ $\sigma _{\gamma h}^{\text{{\rm NLO}
}}(S)/$ $\sigma _{\gamma h}^{\text{{\rm LO}}}(S)$ and $K_{\gamma
h}^{(2)}(S)= $ $\sigma _{\gamma h}^{\text{{\rm NNLO}}}(S)/$ $\sigma _{\gamma
h}^{\text{{\rm NLO}}}(S)$. One can see from Figs.4 and 5 that large
soft-gluon corrections to the production cross sections practically (to
within 1-2 percent) do not affect the Born predictions for $A(S)$ at both
NLO and NNLO.  
\begin{figure}
\begin{center}
\begin{tabular}{cc}
\mbox{\epsfig{file=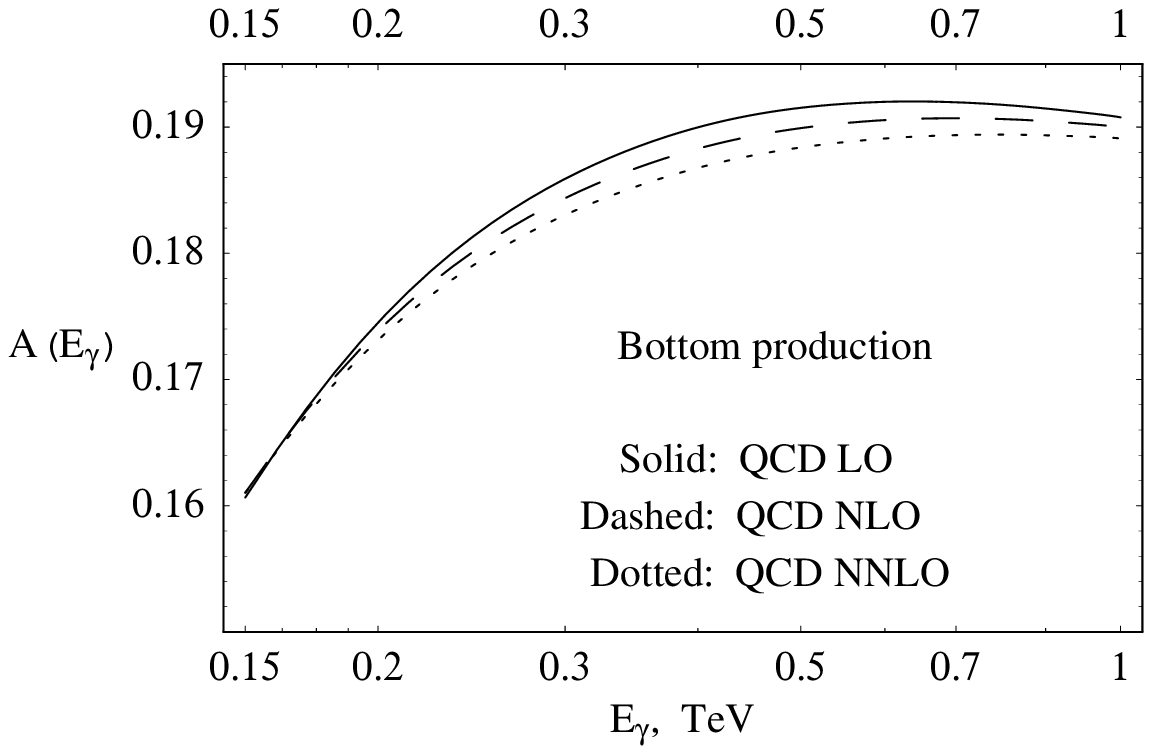,width=250pt}}
&\mbox{\epsfig{file=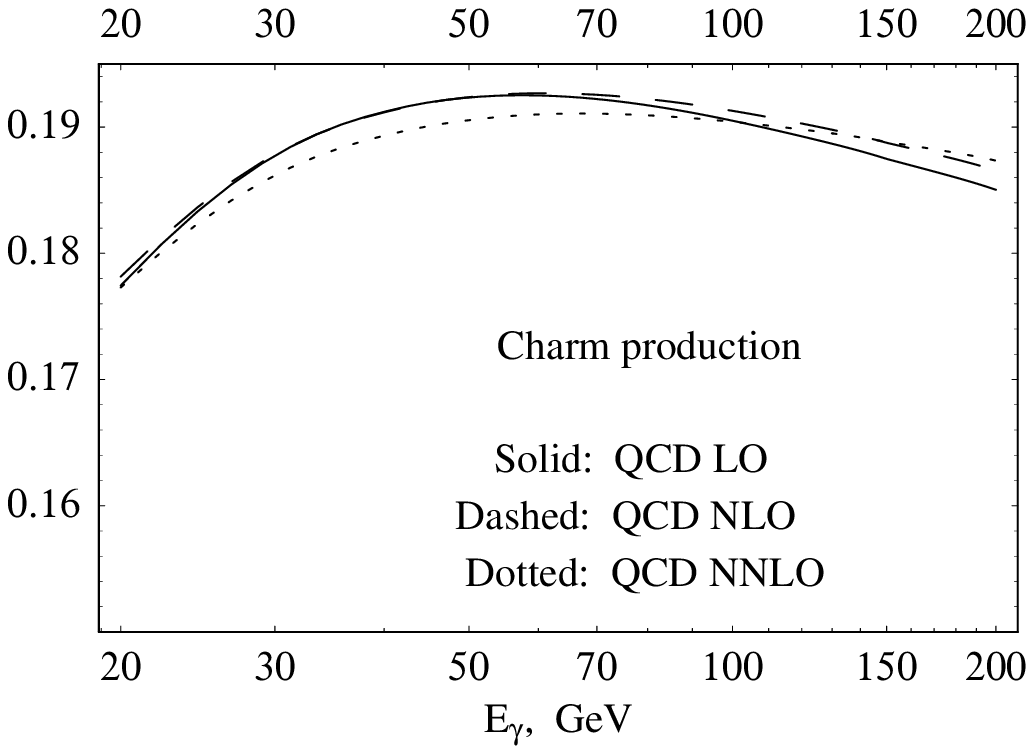,width=250pt}}\\
\end{tabular}
\caption{\small Single spin asymmetry, $A(E_{\gamma})$, in $b$- and $c$-quark
production at LO (solid curve), NLO (dashed curve) and NNLO (dotted curve) to 
NLL accuracy.}
\end{center}
\end{figure}
\begin{figure}
\begin{center}
\begin{tabular}{cc}
\mbox{\epsfig{file=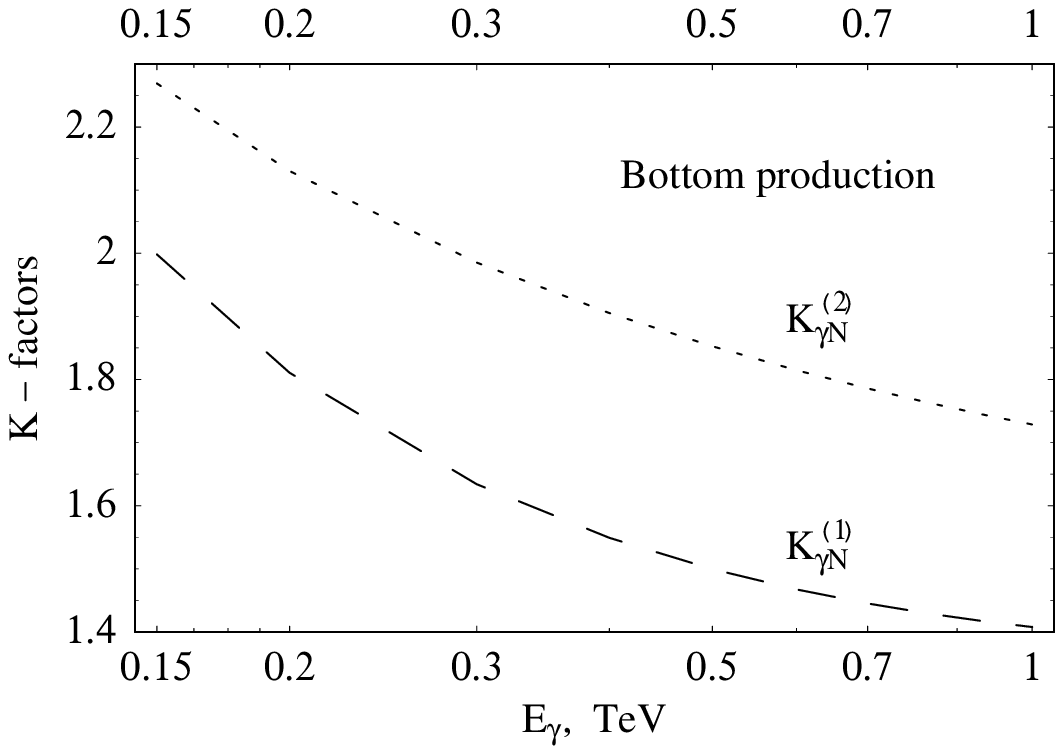,width=250pt}}
&\mbox{\epsfig{file=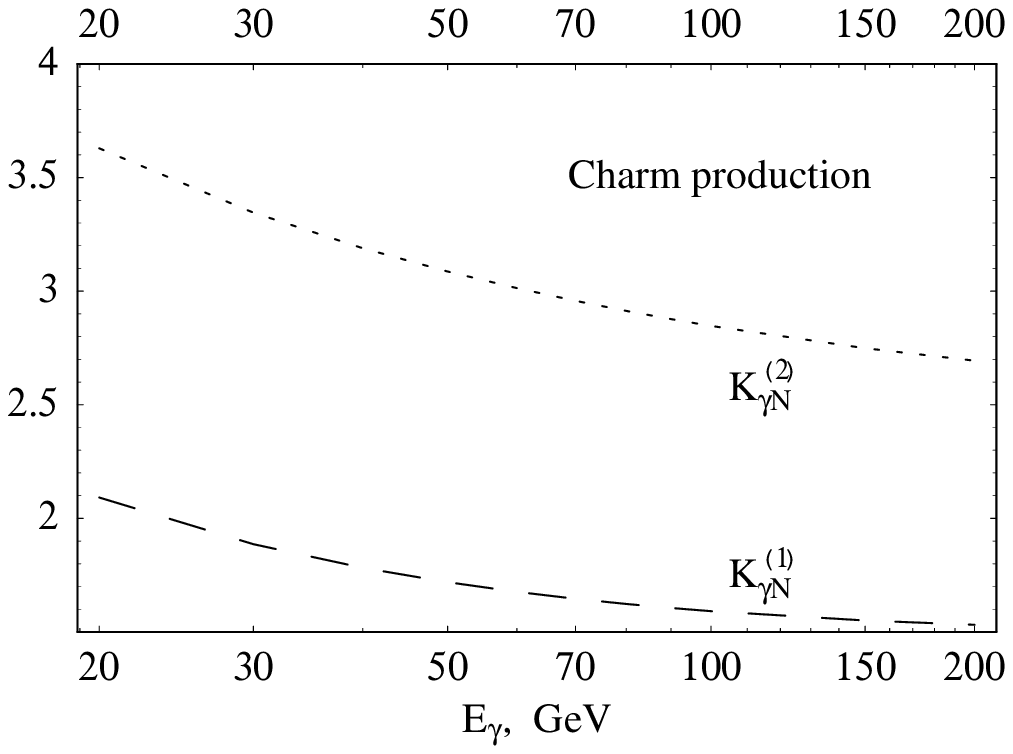,width=250pt}}\\
\end{tabular}
\caption{\small  Energy behavior of the $K(E_{\gamma})$-factors, 
$K_{\gamma h}^{(1)}(S)=$ $\sigma _{\gamma h}^{\text{{\rm NLO}}}(S)/$ 
$\sigma _{\gamma h}^{\text{{\rm LO}}}(S)$ (dashed curve) and $K_{\gamma h}^
{(2)}(S)=$ $\sigma _{\gamma h}^{\text{{\rm NNLO}}}(S)/$ $\sigma _{\gamma h}^
{\text{{\rm NLO}}}(S)$ (dotted curve), in $b$- and $c$-quark production at 
NLL level.}
\end{center}
\end{figure}

Our calculations given in Figs.3-5 represent the central result of this paper. 
At first sight, the situation seems paradoxial: large soft-gluon corrections 
display a strong $\varphi$-dependence in the case of the photon-gluon 
fusion and, simultaneously, are practically $\varphi$-independent at the 
hadron level. A qualitative explanation of this fact will be given in Section IV. 
We shall see that sufficiently soft gluon distribution function leads to a 
factorization of the photon-hadron cross section and, as a consequence, to 
fast convergence of the perturbative series for $A(S)$. 

Another remarkable property of the azimuthal asymmetry closely related 
to fast perturbative convergence is its parametric stability\footnote{Of course, 
parametric stability of the fixed order results does not imply a fast convergence of 
the corresponding series. However, a fast convergent series must be parametricaly 
stable. In particular, it must be $\mu _{R}$- and $\mu _{F}$-independent.}.
As it was shown in \cite{13}, the LO predictions for $A(S)$ are
insensitive (to within few percent) to standard theoretical uncertainties in the 
QCD input parameters: $m$, $\mu _{R}$, $\mu _{F}$, $\Lambda _{QCD}$ and
in the gluon distribution function. We have verified that the same situation 
takes place at higher orders too. In particular, all the CTEQ5 versions of the 
gluon density as well as the MRST parametrizations \cite{mr} lead to asymmetry 
predictions which coincide with each other with accuracy better than 1.5\%. 

Note also the scaling behavior of the azimuthal asymmetry: with a good accuracy
the quantity $A(S)$ is a function of the only variable $\rho _{h}$, 
$\rho_{h}=4m^{2}/S$, so that \cite{13}
\begin{equation}
A(S)\mid _{{\rm {Bottom}}}\approx A\left( Sm_{c}^{2}/m_{b}^{2}\right) \mid _{
{\rm {Charm}}}.  \label{36}
\end{equation} 
This property of the asymmetry reflects its independence from $\Lambda _{QCD}$.

Let us now discuss the hard  ($\vec{p}_{g,T}\ne 0$) and virtual NLO contributions, 
i.e. radiative corrections to the hard functions, 
$H_{\gamma g}(s)$ and $\Delta H_{\gamma g}(s)$.
Since these contributions do not contain the Sudakov logarithms, they are 
expected to be small near the threshold. One can see from Fig.1 that it 
is really the case.  Soft radiation describes very well the exact NLO results on 
the unpolarized photon-gluon fusion at partonic energies up to $\eta\approx  2$. 
Since the gluon distribution function supports just the threshold 
region, the soft-gluon contribution dominates the photon-hadron 
cross section approximately up to $S/4m^2\sim 10$. Using the exact
expression for the $\gamma g$ cross section \cite{25}, we have verified 
that the contribution originating from the region $\eta >2$ makes only 1-2\% 
from the NLO predictions for the unpolarized bottom production  in 
$\gamma p$ collisions at $E_{\gamma }\le$ 1 TeV. In other words, radiative 
corrections to the unpolarized hard function affect the NLL predictions for the 
asymmetry $A(S)$ by about +2\% at $S/4m^2\sim 10$ and by less than 1\% 
at lower energies.

Presently, the exact NLO calculations of the $\varphi$-dependent cross 
section of heavy flavor production are not completed \cite{31}. 
However we can be sure that, at energies not so far from the production 
threshold, the soft radiation is the dominant perturbative mechanism in 
the polarized case too. First, the LO predictions for the $\varphi$-dependent 
cross section are large and the Sudakov logarithms have 
universal, spin-independent structure. For this reason, the polarized heavy 
quark production has also a strong  threshold enhancement. Second, our 
analysis of the exact scale-dependent polarized cross sections given in Fig.2 
confirms with a good accuracy the dominance of the soft-gluon 
contribution. Third, we have verified that soft radiation is a good 
approximation to the threshold behavior of the explicit NLO results \cite{32} 
on  heavy quark production in $\gamma g$ collisions with longitudinally 
polarized initial states too. These facts argue that radiative corrections to 
the $\varphi$-dependent hard function can not also  affect significantly the 
soft-gluon predictions for the asymmetry at the energies under consideration. 

At very  high energies,
$S/4m^2\gtrsim 10^2$, the dominant NLO production mechanism 
is the so-called flavor excitation (FE).  This mechanism arises from the 
diagrams with the $t$-channel gluon exchange and leads to a constant 
value for the unpolarized photon-gluon fusion cross section as 
$s \rightarrow \infty$ \cite{23,25}. In this limit,  the hard component 
associated with the FE mechanism is large and can, in principle, affect the LO 
predictions for the asymmetry. One can expect that  hard corrections will 
result in some dilution of the azimuthal asymmetry at superhigh energies because 
of different asymptotical behavior of the spin-dependent and unpolarized 
cross sections: at $s \rightarrow \infty$,  $ c_{\rm FE}^{(1,0)}(s)\propto const$ 
while ${\Delta c_{\rm FE}^{(1,0)}}(s)\propto \ln s/s$.  
In particular, hard corrections can be sizable in the case of charm 
production at $E_{\gamma }\sim$ 1 TeV. For this reason, we 
consider only the energy region up to $S/4m^2\sim 10$. 

In the conclusion of this section, a remark about the next-to-next-to-leading 
logarithms (NNLL) contribution. In the unpolarized case,  it is possible to estimate 
the contribution of NNLL at NNLO using the method proposed in Refs.\cite{30,12}.
For this purpose, we can add to the NLL approximation (\ref{27}) 
the subleading $\delta \left( s_{4}\right) $-term containing the exact virtual plus soft 
corrections \cite{23,24}. Exponentiating the obtained expression, one can determine 
the coefficients of all the powers of the threshold logarithms 
$[\ln^{k}\left(s_{4}/m^{2}\right)/s_{4}]_{\small +}$ at NNLO. 
(Remember that the NLL expansion (\ref{28}) does not give all the terms with $k=0,1$ 
but only those ones involving the scale $\mu$).
Since the exact expression for $\Delta c^{(1,0)}$ is not presently available, in the 
polarized case we can analyze numerically only the factorized contribution of the so-called 
Coulomb singularity which originates from the final state interaction between the 
massive quarks and dominates in the region very close to the threshold \cite{23}. 
We have verified that the Coulomb corrections to both $\Delta c^{(2,0)}$ and  
$c^{(2,0)}$ are of the order of few percent and that their contribution to 
$A(S)$ is negligible. 

Beyond the NNLO, not all of the subleading logarithms coefficients are under
control in the resummation formalism. At the same time, the NLL predictions for 
partonic cross sections begin to grow rapidly with the order of the perturbative 
expansion in $\alpha_{s}$ due to the renormalon ambiguities \cite{9,10}.
For this reason, the perturbative expansion is usually stopped at NNLO, 
avoiding the theoretical problem with power corrections.
Nevertheless, we shall see in the next section that in the case of single spin 
asymmetry it makes sense to analyze the higher order corrections too. 

\section{Higher order contributions}

\noindent Formally, the finite-order expansion procedure can be extended at NLL 
level to arbitrary higher order in $\alpha _{s}$. According to 
(\ref{17}) and (\ref{18}), the LL contribution can be written as
\begin{equation}
\frac{\text{d}\hat{\sigma}_{\gamma g}^{{\rm LL}}}{\text{d}\varphi }\left(
s,\varphi \right) =\int_{0}^{y_{m}}\text{d}yx_{y}^{3/2}\frac{\text{d}\hat{
\sigma}_{\gamma g}^{{\rm Born}}}{\text{d}\varphi }\left( x_{y}s,\varphi
\right) \left\{ \delta \left( y\right) +\sum_{n=1}^{\infty }\left( \frac{
\alpha _{s}C_{A}}{\pi }\right) ^{n}\frac{2n}{n!}\left[ \frac{\ln ^{2n-1}y}{y}
\right] _{+}\right\} ,  \label{37}
\end{equation}
where $y=s_{4}/m^{2}$, $x_{y}=\left( 1-ym^{2}/s\right) ^{2}$, $y_{m}=(s-2m
\sqrt{s})/m^{2}$ and
\begin{equation}
\frac{\text{d}\hat{\sigma}_{\gamma g}^{{\rm Born}}}{\text{d}\varphi }
(s,\varphi )=\frac{e_{Q}^{2}\alpha _{em}\alpha _{s}}{2s}\left\{ -2\beta
+2\ln \frac{1+\beta }{1-\beta }+\left[ -2\beta \left( 1-\beta ^{2}\right)
+\left( 1-\beta ^{4}\right) \ln \frac{1+\beta }{1-\beta }\right] \left( 1+
{\cal P}_{\gamma }\cos 2\varphi \right) \right\}   \label{38}
\end{equation}
with $\beta =\sqrt{1-4m^{2}/s}$. At hadronic level, we have
\begin{equation}
\Delta \sigma _{\gamma h}^{{\rm LL}}\left( S\right) =\Delta \sigma _{\gamma
h}^{{\rm Born}}\left( S\right) +\sum_{n=1}^{\infty }\left( \frac{\alpha
_{s}C_{A}}{\pi }\right) ^{n}\frac{2n}{n!}A_{2n-1}(S)\sigma _{\gamma
h}^{(2n-1)}\left( S\right) ,  \label{39}
\end{equation}
where
\begin{equation}
(\Delta )\sigma _{\gamma h}^{(i)}\left( S\right) =\frac{1}{S}
\int_{4m^{2}}^{S}\text{d}s\phi _{g/h}(s/S)\int_{0}^{y_{m}}\text{d}
yx_{y}^{3/2}(\Delta )\hat{\sigma}_{\gamma g}^{{\rm Born}}\left(
x_{y}s\right) \left[ \frac{\ln ^{i}y}{y}\right] _{+},  \label{40}
\end{equation}
with $\phi _{g/h}(z)$ the gluon distribution function, and the quantity
\begin{equation}
A_{i}(S)=\frac{\Delta \sigma _{\gamma h}^{(i)}\left( S\right) }{\sigma
_{\gamma h}^{(i)}\left( S\right) }  \label{41}
\end{equation}
describes the partial contribution of the $i$th power of the threshold logarithm.

We have calculated numerically the quantities $A_{i}(S)$ up to $i=11$, i.e., up 
to the 6th order (N$^{6}$LO) of perturbative expansion in $\alpha_{s}$. The results 
of our computations are presented in Fig.6. One can see that $A_{i}(S)$ are close 
to the LO result, $A_{{\rm Born}}(S)=\Delta \sigma _{\gamma h}^{{\rm Born}}
\left(S\right)/\sigma _{\gamma h}^{{\rm Born}}\left( S\right)$, for all $i\leq 11$.
\begin{figure}
\begin{center}
\begin{tabular}{cc}
\mbox{\epsfig{file=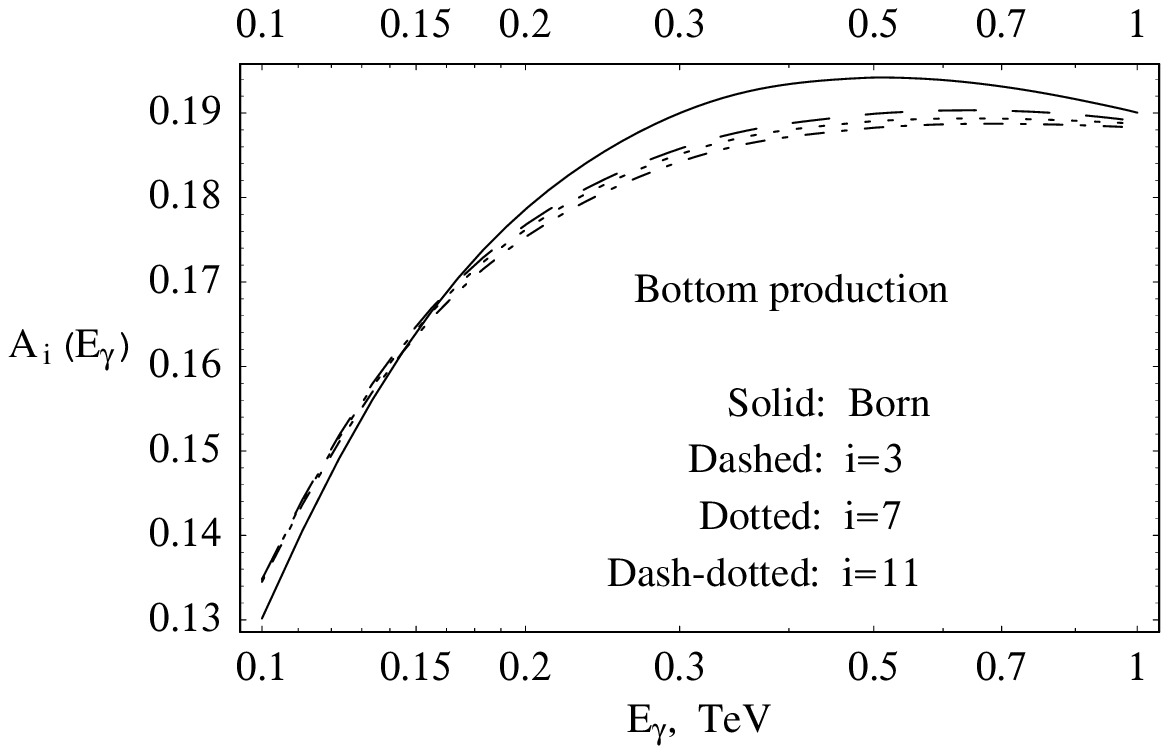,width=250pt}}
&\mbox{\epsfig{file=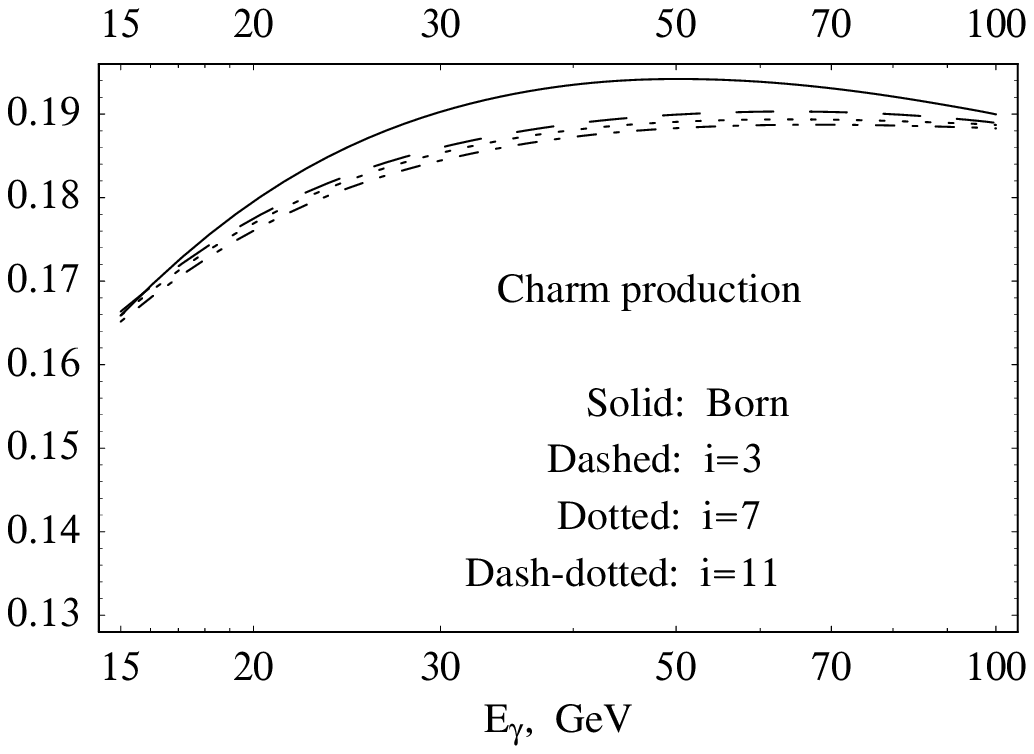,width=250pt}}\\
\end{tabular}
\caption{\small Energy behavior of the $A_{i}(E_{\gamma})$-factors in 
$b$- and $c$-quark production.}
\end{center}
\end{figure}
So, we see that, at higher orders of perturbation theory, the LL predictions 
for the azimuthal asymmetry are also practically insensitive to soft radiation. 
The same situation takes place at the NLL level too.  
Indeed, neglecting in (\ref{17})-(\ref{19}) the small contributions of order 
of $(1-2C_{F}/C_{A})=1/N_{c}^{2}$ and the terms proportional 
to $\ln \left(t_{1}/u_{1}\right)$ which vanish after integration over $t_{1}$ 
and $u_{1}$, we can write
\begin{equation}
\Delta \sigma _{\gamma h}^{{\rm NLL}}\left( S\right) =\sum_{n=1}^{\infty
}\left( \frac{\alpha _{s}C_{A}}{\pi }\right) ^{n}\frac{2n-1}{(n-1)!}\left(
1+(n-1)\frac{2b_{2}}{3C_{A}}\right) A_{2n-2}(S)\sigma _{\gamma
h}^{(2n-2)}\left( S\right)+{\cal O}\left( 1/N_{c}^{2}\right),   \label{42}
\end{equation}
with the partial contributions $A_{i}(S)$ defined by (\ref{41}).
Moreover, one can conclude from Eqs.(\ref{39})-(\ref{41}) and Fig.6 that all 
those subleading logarithmic (SLL) contributions which have the form 
\begin{equation}
(\Delta )\sigma _{\gamma h}^{\rm SLL}\left( S\right) \simeq \sum_{n,i} 
\left(\frac{\alpha_{s}}{\pi }\right)^{n}\frac{C(n,i)}{S}
\int_{4m^{2}}^{S}\text{d}s\phi _{g/h}(s/S)\int_{0}^{y_{m}}\text{d}
yx_{y}^{3/2}(\Delta )\hat{\sigma}_{\gamma g}^{{\rm Born}}\left(
x_{y}s\right) \left[ \frac{\ln ^{i}y}{y}\right] _{+},  \label{43}
\end{equation}
with arbitrary numerical coefficients $C(n,i)$, also can not affect essentially 
the LO predictions for the azimuthal asymmetry.

We have also investigated how the soft-gluon corrections to 
the  asymmetry depend on the choice of the gluon distribution function. 
Parametrizing the gluon density as $z^{-\lambda}(1-z)^{L}$, we 
have variated the parameters $\lambda$ and $L$ in wide intervals: 
$0 \leq \lambda \leq 1$, $0 \leq L \leq 6$. 
The energy behavior of the functions $A_{i}(S)$ was found to be 
different at extreme values of $L$ and $\lambda$ however, at 
$L \ge 2$, the ratio $A_{i}(S)/A_{{\mathrm Born}}(S)$ is equal to unity 
with a good accuracy irrespective of $\lambda$. Moreover, the softer 
the gluon density (i.e., the larger $L$) the smaller the corrections to 
the asymmetry. All curves given in Fig.6 are calculated 
at $\lambda=1$, $L=5$.\footnote{The gluon distribution function 
is presently unknown beyond the
 NLO. We have verified that, in the case of CTEQ5M parametrization, 
 $A_{i}(S)/A_{{\mathrm Born}}(S)$ is equal to unity to within 4\% 
 for all $i\leq 11$.} 

To clarify the origin of perturbative stability of the asymmetry, let us
rewrite the Eqs.(\ref{37})-(\ref{40}) in a more convenient way. Changing the 
order of integrations in (\ref{40}), we obtain
\begin{equation}
(\Delta )\sigma _{\gamma h}^{\mathrm{LL}}\left( S\right) =\sum_{n=0}^{\infty
}\left( \frac{\alpha _{s}C_{A}}{\pi }\right) ^{n}\frac{1}{n!}
\int_{4m^{2}/S}^{1}\text{d}z\phi _{g/h}(z)(\Delta )\hat{\sigma}_{\gamma g}^{
\mathrm{Born}}\left( zS\right) \Phi _{n}(z,S),   \label{44}
\end{equation}
where
\begin{eqnarray}
\Phi _{n}(z,S) &=&-z^{3/2}\phi _{g/h}^{-1}(z)\frac{\partial }{\partial z}
\int_{z}^{1}\frac{\text{d}\tau }{\tau ^{3/2}}\phi _{g/h}(\tau )\ln
^{2n}\left[ \left( \tau -\sqrt{z\tau }\right) S/m^{2}\right]  \\  \nonumber 
&=&\lim_{\epsilon \rightarrow 0}\left[ \ln ^{2n}\frac{\epsilon S}{2m^{2}}
+nz\phi _{g/h}^{-1}(z)\int_{z+\epsilon }^{1}\frac{\text{d}\tau }{\tau }\phi
_{g/h}(\tau )\frac{\ln ^{2n-1}\left[ \left( \tau -\sqrt{z\tau }\right)
S/m^{2}\right] }{\tau -\sqrt{z\tau }}\right]  \\  \label{45}
&=&\ln ^{2n}\frac{\left( 1-\sqrt{z}\right) S}{m^{2}}+2n\int_{0}^{1-\sqrt{z}}
\frac{\text{d}\xi }{\xi }\ln ^{2n-1}\left( \frac{\xi S}{m^{2}}\right) \left[ 
\frac{z\phi _{g/h}\left( ( \sqrt{z}/2+\sqrt{z/4+\xi })
^{2}\right) }{2\phi _{g/h}(z)\left( z/4+\xi +\sqrt{(z/4+\xi )z/4}\right) }
-1\right] . \nonumber 
\end{eqnarray}
Formally, we can write
\begin{equation}
(\Delta )\sigma _{\gamma h}^{\mathrm{LL}}\left( S\right) =\sum_{n=0}^{\infty
}\left( \frac{\alpha _{s}C_{A}}{\pi }\right) ^{n}\frac{1}{n!}\Phi _{n}\left(
(\Delta )z_{n}(S),S\right) \int_{4m^{2}/S}^{1}\text{d}z\phi _{g/h}(z)(\Delta
)\hat{\sigma}_{\gamma g}^{\mathrm{Born}}\left( zS\right) ,  \label{46}
\end{equation}
with the mean value points $4m^{2}/S\leq z_{n}(S), \Delta z_{n}(S)\leq 1$.  

At energies up to $S/4m^2\sim 10$, the functions $\Phi _{n}(z,S)$, $n\ge 1$, 
vary slowly at $z\sim 4m^{2}/S$ and grow at large $z\sim 1$ 
as $\ln ^{2n}\left( 1-\sqrt{z}\right)$. Since the gluon density rapidly vanishes 
at large $z$, $\phi _{g/h}(z)\sim (1-z)^{L}$, both values 
$z_{n}(S)$ and $\Delta z_{n}(S)$ are of the order of $4m^{2}/S$. 
In other words, sufficiently soft gluon distribution function makes the collinear gluon 
radiation effectively soft at hadron level.  In this case, the different high energy behavior 
of the spin-dependent and unpolarized  Born cross sections is irrelevant since 
practically whole contribution to the radiative corrections  to these quantities originates 
from the threshold region. Furthermore, $\hat{\sigma}^{\rm Born}_{\gamma g}(s)$ 
and $\Delta\hat{\sigma}^{\rm Born}_{\gamma g}(s)$ take their maximal values 
practically at the same energies: one can see from Figs.1 and 2 that both Born 
level cross sections have peaks at $\eta\simeq 1$. According to the saddle point 
arguments, these properties of the gluon distribution function and photon-gluon fusion 
provide with a good accuracy the equality\footnote
{Note that Eq.(\ref{47}) does not hold at very high energies 
since the functions $\Phi _{n}(z,S)$ vary strongly in whole interval 
$4m^{2}/S < z < 1$ at $S/4m^2\gtrsim 10^2$. In this case, the different high 
energy behavior of the spin-dependent and unpolarized  Born cross sections 
becomes important and leads to $\varphi$-dependent soft-gluon corrections.}
\begin{equation}
\Phi _{n}\left(\Delta z_{n}(S),S\right)\simeq \Phi _{n}\left(z_{n}(S),S\right),  \label{47}
\end{equation}
which leads to spin-independent radiative factor in (\ref{46}).

As noted in previous Section and shown in Fig.3, soft-gluon corrections to 
the partonic cross section depend essentially on the azimuthal angle $\varphi$. 
So, the mere spin-independent 
structure of the  Sudakov logarithms can not explain our results. Our analysis 
shows that two more factors are responsible for high perturbative stability of the 
hadron level asymmetry. First, at the energies under consideration, 
the gluon distribution function supports the contribution of the threshold region. 
Second, the extremum point of the Born level cross section, $\eta\simeq 1$, 
is practically $\varphi$-independent.

\section{Conclusion}

\noindent In this paper we have investigated the impact of the soft gluon
radiation on the single spin asymmetry in heavy flavor production by linearly
polarized photons.  Our calculations of the spin-dependent cross section 
at the NLL level up to the 6th order in $\alpha_s$ show that the azimuthal 
asymmetry is practically insensitive to the soft-gluon corrections at fixed 
target energies. Taking into account the remarkable properties of $A(S)$ 
observed in our previous paper \cite{13},  we conclude that, 
unlike the unpolarized cross sections, the single spin asymmetry in heavy flavor 
photoproduction is an observable quantitatively well defined in pQCD: it is stable 
both parametricaly and perturbatively, and insensitive to nonperturbative contributions. 
This asymmetry is of leading twist and can be measured at SLAC where the linearly 
polarized photon beam will be available soon \cite{15,16}.
Measurements of the azimuthal asymmetry would provide an ideal test of the pQCD 
applicability to heavy flavor production.
\acknowledgements 
\noindent The author would like to thank A.Capella, Yu.L.Dokshitzer, A.V.Efremov, 
A.B.Kaidalov, G.P.Korchemsky and A.H.Mueller for useful discussions. I am grateful 
to Laboratoire de Physique Theorique, Universite de Paris XI, for hospitality 
while this work has been completed. 
This work was supported in part by the grant NATO PST.CLG.977275.

\end{document}